\documentclass[12pt,a4paper]{article}
\usepackage{eurosym}
\usepackage{mathrsfs}
\usepackage[utf8]{inputenc}
\usepackage[dvipsnames]{xcolor}
\usepackage{dsfont}
\usepackage{amssymb}
\usepackage{amsfonts}
\usepackage{amsmath}
\usepackage{bigints}
\usepackage[bottom]{footmisc}
\usepackage{indentfirst}
\usepackage{endnotes}
\usepackage{graphicx}
\usepackage{rotating}
\usepackage{enumerate}
\usepackage[thmmarks,amsmath,amsthm,hyperref]{ntheorem}
\usepackage{enumitem}
\usepackage{hyperref}
\usepackage[round]{natbib}
\usepackage{multirow}
\usepackage[flushleft]{threeparttable}
\usepackage{booktabs}
\usepackage{lastpage}
\usepackage{amsfonts}
\usepackage{amsmath}
\usepackage[a4paper,margin=1in]{geometry}
\usepackage[nodisplayskipstretch]{setspace}
\usepackage{graphicx}
\usepackage{float}
\usepackage{color}
\usepackage{subcaption}
\usepackage{multirow}
\usepackage{pgffor}
\usepackage{mathtools}
\usepackage{amssymb}
\usepackage{eurosym}
\usepackage{natbib}
\usepackage{hyperref}
\usepackage{mathtools}
\usepackage{amsfonts}
\usepackage{eurosym}
\usepackage{amssymb}
\usepackage{amsmath}
\usepackage{graphicx}
\usepackage{float}
\usepackage{color}
\usepackage{natbib}
\usepackage{hyperref}
\usepackage[nodisplayskipstretch]{setspace}
\usepackage{multirow}
\usepackage{subcaption}
\usepackage{pgffor}
\usepackage{booktabs}
\usepackage{adjustbox}
\usepackage{graphicx}
\usepackage{grffile}
\usepackage[table]{xcolor}

\definecolor{rowgray}{gray}{0.95}
\theoremstyle{definition}
\newtheorem{theorem}{Theorem}
\newtheorem{assumption}{Assumption}

\newtheorem{lemma}[theorem]{Lemma}

\newtheorem{proposition}{Proposition}

\theoremstyle{break}
\makeatletter
\def\@biblabel#1{\hspace*{-\labelsep}}
\makeatother
\newdimen\dummy
\dummy=\oddsidemargin
\input{tcilatex}
\begin{document}

\title{Confidence Sets under Weak Identification: Theory and Practice\thanks{
This paper is based on the MA dissertation of Gustavo Schlemper, written
under the supervision of Marcelo J. Moreira. The manuscript substantially
builds on that dissertation and is the result of a collaboration between the
two authors. We are especially grateful to Mahrad Sharifvaghefi for lengthy
discussions that significantly shaped the development of this project. We
thank Marinho Bertanha, Luan Borelli, Guilherme Exel, Marcelo Fernandes,
Jack Porter, and Miguel Troppmair for helpful comments, and Pedro Watuh\~{a}
for excellent research assistance. This study was financed in part by CAPES
(Finance Code 001), CNPq, and FAPERJ. Emails: mjmoreira@fgv.br and
gschlemp@stanford.edu.}}
\author{Marcelo J. Moreira \\
\emph{FGV EPGE} \and Gustavo Schlemper \\
\emph{Stanford University} }
\date{\today}
\maketitle

\begin{abstract}
We develop new methods for constructing confidence sets and intervals in
linear instrumental variables (IV) models based on tests that remain valid
under weak identification and under heteroskedastic, autocorrelated, or
clustered errors. In practice, researchers typically recover such sets by
grid search, a procedure that can miss parts of the confidence region,
truncate unbounded sets, and deliver misleading inference. We replace grid
inversion with exact and approximation-based methods that are both reliable
and computationally efficient.

Our approach exploits the polynomial and rational structure of the
Anderson-Rubin and Lagrange multiplier statistics to obtain exact confidence
sets via polynomial root finding. For the conditional quasi-likelihood ratio
test, we derive an exact inversion algorithm based on the geometry of the
statistic and its critical value function. For more general conditional
tests, we construct polynomial approximations whose coverage error vanishes
with approximation degree, allowing numerical accuracy to be made
arbitrarily high. In many empirical applications with weak instruments,
standard grid methods produce incorrect confidence regions, while our
procedures reliably recover sets with correct nominal coverage.

The framework extends beyond linear IV to models with piecewise polynomial
or rational moment conditions, offering a general tool for reliable
weak-identification robust inference.
\end{abstract}

\bigskip \newpage

\section{Introduction}

Weak instruments can distort statistical inference in instrumental variables
(IV) models. When identification is weak, conventional $t$-based confidence
intervals may have zero asymptotic coverage probability. \citet{Dufour97}
shows that any confidence interval that is bounded with probability one,
such as the usual interval $[\hat{\beta}\pm 1.96\cdot \text{se}(\hat{\beta}%
)] $, has zero confidence level under weak identification. In practical
terms, intervals that appear informative may fail to contain the true
structural parameter altogether. Reliable inference therefore requires tests
that control size regardless of instrument strength, together with
confidence sets obtained by inverting such tests.

A central but often overlooked issue is that even when valid tests are
available, empirical inference depends on how these tests are inverted.
Standard numerical procedures used in applied work can produce incorrect
confidence sets. In particular, commonly used grid-search methods may
include parameter values that are rejected by the underlying test and
exclude values that should belong to the confidence set. As a result,
empirical conclusions can depend on the numerical inversion method rather
than on the statistical procedure itself.

Consider the linear IV model
\begin{eqnarray}
y_{1i} &=& y_{2i}\beta + x_{i}^{\prime}\gamma_{1} + u_{i},  \label{eq:model}
\\
y_{2i} &=& z_{i}^{\prime}\pi + x_{i}^{\prime}\gamma_{2} + v_{2i},  \notag
\end{eqnarray}
where $y_{1i}$ is the outcome, $y_{2i}$ is the endogenous regressor of
interest, $z_{i}$ is a vector of instruments, and $x_{i}$ contains exogenous
covariates. A leading example is the Euler equation design of \citet{Yogo04}%
, where $y_{1i}$ represents consumption growth, $y_{2i}$ is an asset return
or interest rate, and $\beta$ corresponds to the elasticity of intertemporal
substitution. The instrument vector consists of lagged financial predictors
such as interest rates, dividend--price ratios, and inflation. In this
environment the instruments forecast returns only weakly, making
conventional inference unreliable.

A large literature has developed tests that remain valid under weak
identification. In the classical homoskedastic IV model with one endogenous
regressor and $k$ instruments, \citet{AndersonRubin49} introduce the AR test
with a $\chi_k^2$ distribution that is robust to weak instruments. %
\citet{Moreira02} shows that this test is optimal in the just-identified
case and that additional instruments can improve power. Building on this
idea, he demonstrates that several tests, including a score (LM) test, are
robust to weak identification. \citet{Moreira03} proposes a conditional
approach that replaces fixed critical values with conditional quantiles and
introduces the conditional likelihood ratio (CLR) test. %
\citet{AndrewsMoreiraStock06} show that the CLR test satisfies natural
invariance properties and is nearly optimal. These advances establish valid
tests, but they do not resolve a key practical issue: how to reliably invert
these tests to obtain confidence sets.

In parallel with the development of these tests, researchers have studied
the geometry and computation of the associated confidence sets. Although
confidence sets need not be intervals, associated confidence intervals can
be defined as the smallest interval that contains the set (convex hull). %
\citet{DufourTaamouti05} highlight the quadratic structure of AR and LM
confidence sets and provide conditions for boundedness and projection
methods for restricted parameters. \citet{Mikusheva10} introduces an
algorithm for inverting the CLR test under homoskedasticity. These results
show that exact inversion is possible in important special cases, but they
do not provide general procedures applicable in the settings most commonly
encountered in empirical work.

Most empirical applications allow for heteroskedasticity, autocorrelation,
or clustering. Extensions of the AR, LM, and CLR tests to general HAC
settings have been developed by \citet{StockWright00}, %
\citet{AndrewsMoreiraStock04}, \citet{Kleibergen05}, %
\citet{AndrewsMikusheva16}, and \citet{MoreiraMoreira19}, among others. %
\citet{MoreiraRidderSharifvaghefi25} also proposes the conditional
integrated likelihood (CIL) test, which has a Bayesian interpretation. These
results establish valid tests under weak identification and general error
structures, making weak-instrument-robust inference feasible in a wide range
of empirical environments.

However, validity of the test does not guarantee reliable inversion. Outside
the homoskedastic case, both the test statistics and their critical value
functions become complicated functions of the structural parameter. The
algebra that allows closed-form inversion in simple settings no longer
applies directly, and constructing the confidence set becomes a problem that
can materially distort inference if handled poorly.

In practice, researchers typically resort to grid search. This step is often
treated as a routine implementation detail, but it can materially affect the
resulting inference. Using 158 empirical specifications from five well-known
IV applications in macroeconomics, labor, and public finance, we show that
these failures are not rare. Grid inversion frequently fails to recover the
true confidence set and often produces materially different regions relative
to exact inversion. In a substantial fraction of cases, grid methods miss
disconnected components or fail to detect sharp features of the confidence
set. In others, they distort its qualitative shape, for example by reporting
bounded intervals when the true confidence set is unbounded, or by reporting
nonempty intervals when the exact procedure yields an empty set. As a
result, reported confidence intervals may include values that are rejected
by the underlying test or omit values that belong to the confidence set.
These discrepancies can be economically large and arise solely from the
numerical inversion step rather than from the underlying statistical
procedure. In contrast, our methods recover the full confidence set with
correct nominal coverage under general HAC errors.

We develop new methods for constructing confidence sets and intervals for
weak-identification robust tests. For the AR and LM tests, we exploit the
fact that the test statistics are rational polynomial functions of the
structural parameter to obtain exact confidence sets by solving for all
polynomial roots. For the conditional quasi-likelihood ratio (CQLR) test, we
use the statistic's monotonicity and convexity properties to derive an exact
inversion algorithm.

For more general conditional tests, including CLR and CIL, we propose a
simple approximation procedure with two steps. First, we compactify the
parameter space to a bounded interval and evaluate the test at Chebyshev
nodes, which allows for uniform control of the approximation error. Second,
we approximate the test procedure as a function of the hypothesized
parameter by a polynomial whose degree controls numerical accuracy.\footnote{%
Chebyshev approximations are widely used in economics. For example, %
\citet{Heckman74} employ Chebyshev--Hermite expansions in structural
labor-supply estimation, \citet{RennerSchmedders15} use Chebyshev
polynomials to transform non-polynomial expected-utility problems into
polynomial optimization problems, and \citet{TaylorUhlig16} survey
projection methods in macroeconomics built on Chebyshev approximations.}
Confidence sets are then obtained by solving a polynomial inequality, which
allows us to recover all components of the set, including unbounded regions,
without relying on a grid.

We compare three numerical approaches: evenly spaced grids commonly used in
empirical work, such as those implemented by the \texttt{weakiv} command in
Stata; grids based on Chebyshev nodes; and our Chebyshev approximation
method. Changing the grid alone improves node placement, but it is not
enough to recover the full confidence set reliably. The approximation step
is essential.

Across 158 empirical specifications from five well-known IV applications,
standard grid procedures frequently fail. For example, the \texttt{weakiv}
command can produce qualitatively incorrect inference by misclassifying
whether the confidence set is bounded or unbounded. Such failures occur in
more than 40\% of specifications. Our methods recover confidence sets with
correct nominal coverage. Numerical inversion is therefore not a secondary
implementation detail, but a central component of valid weak-identification
robust inference in practice.

Although the paper focuses on the linear IV model, the same algebraic
structure applies to any model with piecewise polynomial or rational moment
conditions. Our methodology therefore extends beyond IV to a broader class
of econometric models with nonlinear identification features.

The remainder of the paper proceeds as follows. Section \ref%
{section:emp_relevance} illustrates the empirical importance of reliable
inversion. Section \ref{section:applications} compares grid search with our
methods. Section \ref{section:derivation} develops exact and approximate
inversion algorithms. Section \ref{section:RF model} presents theoretical
details. Section \ref{section:extensions} extends the framework to more
general models. Section \ref{section:conclusion} concludes.

\section{Current Practice: An Example}

\label{section:emp_relevance}

We illustrate the empirical relevance of our methods using the Euler
equation application of \citet{Yogo04}, a leading example in applied work on
the equity premium. This setting highlights why confidence sets based on
tests that are robust to weak identification and HAC errors are essential.
It also reveals the limitations of standard grid-search procedures, which
can miss components of the confidence set and produce intervals that are too
wide or even qualitatively incorrect. In some cases, grid methods even fail
to detect that the confidence set is empty or unbounded.

The Euler equation implies the linear relationship%
\begin{equation}
r_{i,t+1}=\mu _{i}+\dfrac{1}{\psi }\Delta c_{t+1}+\eta _{i,t+1}
\label{eq:EIS_r_c}
\end{equation}%
where $\Delta c_{t}$ is the log of the consumption at time $t$, $%
r_{i,t}=\log (1+R_{i,t})$, $R_{i,t}$ is the gross real return of asset $i$
at time $t$, and $\psi $ is the elasticity of intertemporal substitution.
Under power utility, it represents the inverse of the relative risk
aversion. Up to a linear transformation, Equation \ref{eq:EIS_r_c} is
equivalent to
\begin{equation}
\Delta c_{t+1}=\gamma _{i}+\psi r_{i,t+1}+\xi _{i,t+1}  \label{eq:EIS_c_r}
\end{equation}%
\citet{Yogo04} uses four instruments: the twice-lagged nominal interest
rate, inflation, consumption growth, and the log dividend-price ratio.

\subsection{Weak Identification and Heteroskedastic Errors in the Euler
Equation}

The Euler equation application of \citet{Yogo04} provides a canonical
setting where weak identification and heteroskedastic errors arise
simultaneously. The instruments have limited predictive power for
consumption growth and asset returns, so identification is often weak, while
macroeconomic data are heteroskedastic. This combination makes it a
particularly demanding environment for inference and a natural setting to
evaluate procedures that claim robustness to both features.

Inference procedures differ sharply along these two dimensions. The
conventional t ratio interval under homoskedasticity is robust to neither
weak identification nor heteroskedasticity. The heteroskedasticity-robust t
ratio corrects only the second dimension, but remains invalid under weak
identification. In contrast, the AR, LM, CQLR, CLR, and CIL tests are
constructed to control size under weak identification. When implemented
under homoskedasticity, they are robust to weak instruments, but not to
heteroskedasticity. Their heteroskedasticity-robust versions correct both
dimensions simultaneously.

\citet{Yogo04} reports confidence intervals rather than general confidence
sets. To facilitate comparison, we also report confidence intervals. These
intervals are obtained by inverting the corresponding tests using our
methods. For the AR, LM, and CQLR tests, inversion is exact. For the CLR and
CIL tests, which do not admit an exact algebraic characterization of the
acceptance region, we compute approximate intervals using the
Chebyshev-based procedures developed in this paper. Although the derivations
rely on the algebraic structure of the test statistics, the final output is
conventional: a confidence interval or confidence set that can be reported
and interpreted in the usual way.

Table \ref{tab:EIS_interest} reports confidence intervals for the elasticity
of intertemporal substitution in Equation \eqref{eq:EIS_c_r}, using interest
rates as the endogenous regressor, for eleven developed countries: Australia
(AUL), Canada (CAN), France (FRA), Germany (GER), Italy (ITA), Japan (JAP),
Netherlands (NTH), Sweden (SWE), Switzerland (SWI), United Kingdom (UK), and
United States (USA). The countries discussed explicitly in the text are
highlighted with gray shading in the table.

In Section~\ref{supp:emp_relevance} in the appendix, we repeat the analysis
using stock returns rather than interest rates as the endogenous variable.
Because stock returns are less predictable, weak identification is more
severe in these specifications. As a result, weak-IV robust procedures often
yield unbounded confidence sets.

\begin{table}[h]
\scalebox{0.862}{ \centering
\begin{threeparttable}
\caption{Confidence Intervals for the EIS Using Interest Rates}
\label{tab:EIS_interest}
\renewcommand{\arraystretch}{1.12}
\begin{tabular}{cccccccc}
\toprule
 & Robust & t-ratio & AR & LM & CQLR & CLR & CIL \\ \midrule
AUL & Yes & {[}-0.19, 0.28{]} & {[}-0.11, 0.22{]} & {[}$-\infty, +\infty${]} & {[}-0.16, 0.28{]} & {[}-0.18, 0.28{]} & {[}-0.20, 0.31{]} \\
AUL & No & {[}-0.17, 0.26{]} & {[}-0.14, 0.20{]} & {[}-0.22,
13.48{]} & {[}-0.21, 0.26{]} & {[}-0.21, 0.26{]} & {[}-0.15, 0.30{]}
\\ \midrule \rowcolor{rowgray}
CAN & Yes & {[}-0.64, 0.03{]} & {[}-0.55, -0.16{]} & {[}-0.85, 250.88{]} & {[}-0.82, 0.09{]} & {[}-0.80, 0.07{]} & {[}-0.77, 0.04{]} \\
\rowcolor{rowgray} CAN & No & {[}-0.61, -0.00{]} & {[}-0.51,
-0.17{]} & {[}-0.72, 13.74{]} & {[}-0.70, -0.01{]} & {[}-0.70,
-0.01{]} & {[}-0.66, 0.08{]} \\ \midrule
FRA & Yes & {[}-0.38, 0.22{]} & {[}-0.56, 0.31{]} & {[}-45.23, 0.16{]} & {[}-0.39, 0.16{]} & {[}-0.40, 0.16{]} & {[}-0.41, 0.15{]} \\
FRA & No & {[}-0.46, 0.29{]} & {[}-0.66, 0.52{]} & {[}-49.85,
0.30{]} & {[}-0.46, 0.31{]} & {[}-0.46, 0.31{]} & {[}-2.36, 2.15{]}
\\ \midrule \rowcolor{rowgray}
GER & Yes & {[}-1.45, 0.61{]} & {[}-1.73, 0.66{]} & {[}-110.06, 0.34{]} & {[}-1.38, 0.34{]} & {[}-1.38, 0.36{]} & {[}-1.30, 0.41{]} \\
\rowcolor{rowgray} GER & No & {[}-1.09, 0.25{]} & {[}-1.52, 0.50{]}
& {[}-1.18, 15.91{]} & {[}-1.19, 0.24{]} & {[}-1.18, 0.24{]} &
{[}-1.10, 1.20{]} \\ \midrule
ITA & Yes & {[}-0.23, 0.09{]} & {[}-0.29, 0.18{]} & {[}-4.85, 0.10{]} & {[}-0.23, 0.11{]} & {[}-0.23, 0.10{]} & {[}-0.24, 0.11{]} \\
ITA & No & {[}-0.23, 0.09{]} & {[}-0.29, 0.17{]} & {[}-6.45, 0.11{]}
& {[}-0.23, 0.11{]} & {[}-0.23, 0.11{]} & {[}-0.25, 0.25{]} \\
\midrule
JAP & Yes & {[}-0.46, 0.38{]} & {[}-0.88, 0.25{]} & {[}$-\infty, +\infty${]} & {[}-0.77, 0.20{]} & {[}-0.82, 0.19{]} & {[}-0.84, 0.18{]} \\
JAP & No & {[}-0.44, 0.37{]} & {[}-0.57, 0.46{]} & {[}$-\infty,
+\infty${]} & {[}-0.55, 0.44{]} & {[}-0.54, 0.44{]} & {[}-1.58,
0.30{]} \\ \midrule
NTH & Yes & {[}-0.65, 0.35{]} & $\emptyset$ & {[}$-\infty, +\infty${]} & {[}-0.54, 0.22{]} & {[}-0.56, 0.26{]} & {[}-0.56, 0.28{]} \\
NTH & No & {[}-0.68, 0.38{]} & {[}-0.87, 0.60{]} & {[}$-\infty,
+\infty${]} & {[}-0.73, 0.46{]} & {[}-0.73, 0.46{]} & {[}-2.84,
2.46{]} \\ \midrule
SWE & Yes & {[}-0.19, 0.19{]} & {[}-0.26, 0.26{]} & {[}$-\infty, +\infty${]} & {[}-0.19, 0.19{]} & {[}-0.19, 0.19{]} & {[}-0.20, 0.18{]} \\
SWE & No & {[}-0.19, 0.19{]} & {[}-0.29, 0.28{]} & {[}$-\infty,
+\infty${]} & {[}-0.21, 0.20{]} & {[}-0.21, 0.20{]} & {[}-0.22,
0.24{]} \\ \midrule
SWI & Yes & {[}-1.03, 0.06{]} & {[}-1.33, 0.26{]} & {[}-1.03, 5.89{]} & {[}-1.03, 0.05{]} & {[}-0.99, 0.06{]} & {[}-1.01, 0.06{]} \\
SWI & No & {[}-1.05, 0.07{]} & {[}-1.63, 0.34{]} & {[}-1.17, 7.44{]}
& {[}-1.20, 0.07{]} & {[}-1.18, 0.06{]} & {[}-1.37, 0.11{]} \\
\midrule \rowcolor{rowgray}
UK & Yes & {[}-0.07, 0.40{]} & {[}0.19, 0.28{]} & {[}-0.95, 8.16{]} & {[}-0.68, 9.45{]} & {[}-0.17, 0.48{]} & {[}-0.19, 0.45{]} \\
\rowcolor{rowgray} UK & No & {[}-0.07, 0.40{]} & {[}0.07, 0.25{]} &
{[}$-\infty, +\infty${]} & {[}-0.11, 0.42{]} & {[}-0.11, 0.42{]} &
{[}-1.21, 1.01{]} \\ \midrule \rowcolor{rowgray}
USA & Yes & {[}-0.09, 0.20{]} & $\emptyset$ & {[}$-\infty, +\infty${]} & {[}-0.23, 0.11{]} & {[}-0.27, 0.12{]} & {[}-0.36, 0.15{]} \\
\rowcolor{rowgray} USA & No & {[}-0.11, 0.23{]} & $\emptyset$ &
{[}$-\infty, +\infty${]} & {[}-0.22, 0.23{]} & {[}-0.22, 0.22{]} &
{[}-4.88, 0.86{]} \\ \bottomrule
\end{tabular}

\begin{tablenotes}\footnotesize
\item Column "Robust" indicates whether the CS is robust to heteroskedasticity or not. $\emptyset$ indicates an empty CS.
\item CIs are calculated using our methodology
\end{tablenotes}
\end{threeparttable}
}
\end{table}

Table \ref{tab:EIS_interest} shows that allowing for heteroskedasticity can
materially alter inference. The key pattern is that robust procedures can
substantially change both the location and the shape of the confidence sets,
sometimes dramatically.

Germany provides a clear example. The heteroskedasticity-robust t ratio
interval shifts from [-1.09, 0.25] to [-1.45, 0.61], indicating a sizable
change in uncertainty. The effect is even more pronounced for weak-IV robust
procedures. The LM interval changes from [-1.18, 15.91] to [-110.06, 0.34],
reflecting a drastic reallocation of mass toward extreme negative values and
a sharp contraction on the upper end. Similar, though less extreme, shifts
occur for the AR and CQLR intervals. Among conditional procedures, the
effects are more nuanced. The CLR interval widens under heteroskedasticity,
whereas the CIL interval changes asymmetrically: its lower bound shifts
outward while its upper bound contracts. These differences arise solely from
the choice of covariance estimator, highlighting that inference can be
highly sensitive to how sampling uncertainty is modeled in weakly identified
settings.

The contrast between t ratio intervals and weak IV robust intervals is also
pronounced. For the United States the heteroskedasticity-robust t ratio
interval ranges from -0.09 to 0.20 and is bounded and nonempty. In contrast,
the heteroskedasticity-robust AR interval is empty and the LM interval is
unbounded. In particular, the t ratio procedure never signals lack of
identification, since it always produces a bounded interval.

Even among weak IV robust procedures, behavior differs. For Canada the
heteroskedasticity-robust LM interval ranges from -0.85 to 250.88, much
wider than the corresponding CLR and CIL intervals. For the United Kingdom
the heteroskedasticity-robust CQLR interval ranges from -0.68 to 9.45, again
much wider than the CLR and CIL intervals.

We do not take a normative stand on which test researchers should adopt.
Each procedure embodies a different tradeoff between robustness, power, and
computational complexity. The purpose of this example is to illustrate that,
once weak identification and heteroskedastic errors are taken seriously,
inference can differ substantially across procedures even in familiar
empirical applications. Our contribution is to compute these objects exactly
or with controlled approximation error, ensuring that the reported
confidence regions faithfully reflect the underlying test.

That said, existing theory provides guidance in over-identified settings. It
is well-known that the AR test can be inefficient when there is more than
one instrument (\citet{Moreira02, Moreira09a}). Recent work also documents
non-trivial power losses for LM and CQLR procedures in over-identified
models (\citet{MoreiraRidderSharifvaghefi23}). Conditional procedures such
as CLR and CIL are designed to address these efficiency concerns while
preserving size control under weak identification. For this reason, in
empirical applications with multiple instruments, CLR and CIL often provide
an attractive balance between robustness and precision. Our methods make
their implementation as reliable as that of AR and LM.

In Section 2.2 we compare these intervals with those obtained using grid
search in the same application. The eleven country specifications of %
\citet{Yogo04} provide a transparent setting in which we can directly
contrast the confidence intervals produced by our exact and controlled
approximation methods with those reported by \citet{Yogo04} using grid
search.

\subsection{The Problem with Grid Search Inversion}

In applied work, confidence sets are often obtained by evaluating a test
statistic on a grid over the parameter space and collecting the parameter
values that are not rejected. This approach requires two arbitrary choices:
a compact interval over which the parameter is searched and the number of
grid points used within that interval. In the linear IV model, the parameter
space is the entire real line. Any finite grid therefore imposes truncation,
and its resolution determines whether narrow components of the confidence
set are detected.

Table \ref{tab:compare_yogo} compares the confidence intervals reported by %
\citet{Yogo04}, which are based on grid inversion, with those obtained using
our exact and controlled approximation methods. The comparison is conducted
for the same eleven country specifications analyzed in Section 2.1. The
countries discussed explicitly in the text are highlighted with gray shading
in the table.

At the time \citet{Yogo04} was published, heteroskedasticity-robust
implementations were available for the AR test, but the LM and CLR
procedures were typically implemented under homoskedasticity. For this
reason, comparisons involving heteroskedasticity-robust procedures focus on
the AR test, while comparisons for LM and CLR are conducted under
homoskedasticity to match the specification used by \citet{Yogo04}.

\begin{table}[h]
\scalebox{0.85}{ \centering
\begin{threeparttable}
\caption{Comparison of Yogo's Reports and Our Method}
\label{tab:compare_yogo}
\renewcommand{\arraystretch}{1.12}
\begin{tabular}{cccccccc}
\toprule
 & Robust & Yogo - AR & AR & Yogo - LM & LM & Yogo - CLR & CLR \\ \midrule
\rowcolor{rowgray}
AUL & Yes & {[}-0.17, 0.30{]} & {[}-0.11, 0.22{]} &  & {[}$-\infty, +\infty${]} &  & {[}-0.18, 0.28{]} \\
\rowcolor{rowgray} AUL & No & {[}-0.16, 0.21{]} & {[}-0.14, 0.20{]}
& {[}-0.22, 13.74{]} & {[}-0.22, 13.48{]} & {[}-0.22, 0.27{]} &
{[}-0.21, 0.26{]} \\ \midrule
CAN & Yes & {[}-0.77, 0.11{]} & {[}-0.55, -0.16{]} &  & {[}-0.85, 250.88{]} &  & {[}-0.80, 0.07{]} \\
CAN & No & {[}-0.54, -0.14{]} & {[}-0.51, -0.17{]} & {[}-0.73,
14.15{]} & {[}-0.72, 13.74{]} & {[}-0.71, 0.00{]} & {[}-0.70,
-0.01{]} \\ \midrule
FRA & Yes & {[}-0.57, 0.36{]} & {[}-0.56, 0.31{]} &  & {[}-45.23, 0.16{]} &  & {[}-0.40, 0.16{]} \\
FRA & No & {[}-0.68, 0.53{]} & {[}-0.66, 0.52{]} & {[}-0.47, 0.31{]}
& {[}-49.85, 0.30{]} & {[}-0.48, 0.33{]} & {[}-0.46, 0.31{]} \\
\midrule \rowcolor{rowgray}
GER & Yes & {[}-1.95, 1.63{]} & {[}-1.73, 0.66{]} &  & {[}-110.06, 0.34{]} &  & {[}-1.38, 0.36{]} \\
\rowcolor{rowgray} GER & No & {[}-1.57, 0.54{]} & {[}-1.52, 0.50{]}
& {[}-1.21, 0.26{]} & {[}-1.18, 15.91{]} & {[}-1.23, 0.28{]} &
{[}-1.18, 0.24{]} \\ \midrule
ITA & Yes & {[}-0.34, 0.20{]} & {[}-0.29, 0.18{]} &  & {[}-4.85, 0.10{]} &  & {[}-0.23, 0.10{]} \\
ITA & No & {[}-0.29, 0.18{]} & {[}-0.29, 0.17{]} & {[}-0.24, 0.11{]}
& {[}-6.45, 0.11{]} & {[}-0.24, 0.12{]} & {[}-0.23, 0.11{]} \\
\midrule
JAP & Yes & {[}-0.93, 0.39{]} & {[}-0.88, 0.25{]} &  & {[}$-\infty, +\infty${]} &  & {[}-0.82, 0.19{]} \\
JAP & No & {[}-0.60, 0.49{]} & {[}-0.57, 0.46{]} & {[}$-\infty,
+\infty${]} & {[}$-\infty, +\infty${]} & {[}-0.56, 0.45{]} &
{[}-0.54, 0.44{]} \\ \midrule
NTH & Yes & {[}-0.57, 0.09{]} & $\emptyset$ &  & {[}$-\infty, +\infty${]} &  & {[}-0.56, 0.26{]} \\
NTH & No & {[}-0.91, 0.64{]} & {[}-0.87, 0.60{]} & {[}$-\infty,
+\infty${]} & {[}$-\infty, +\infty${]} & {[}-0.76, 0.48{]} &
{[}-0.73, 0.46{]} \\ \midrule
SWE & Yes & {[}-0.28, 0.28{]} & {[}-0.26, 0.26{]} &  & {[}$-\infty, +\infty${]} &  & {[}-0.19, 0.19{]} \\
SWE & No & {[}-0.30, 0.29{]} & {[}-0.29, 0.28{]} & {[}$-\infty,
+\infty${]} & {[}$-\infty, +\infty${]} & {[}-0.22, 0.21{]} &
{[}-0.21, 0.20{]} \\ \midrule
SWI & Yes & {[}-1.42, 0.50{]} & {[}-1.33, 0.26{]} &  & {[}-1.03, 5.89{]} &  & {[}-0.99, 0.06{]} \\
SWI & No & {[}-1.69, 0.37{]} & {[}-1.63, 0.34{]} & {[}-1.19, 0.07{]}
& {[}-1.17, 7.44{]} & {[}-1.22, 0.09{]} & {[}-1.18, 0.06{]} \\
\midrule \rowcolor{rowgray}
UK & Yes & {[}-0.45, 0.51{]} & {[}0.19, 0.28{]} &  & {[}-0.95, 8.16{]} &  & {[}-0.17, 0.48{]} \\
\rowcolor{rowgray} UK & No & {[}0.04, 0.28{]} & {[}0.07, 0.25{]} &
{[}$-\infty, +\infty${]} & {[}$-\infty, +\infty${]} & {[}-0.12,
0.43{]} & {[}-0.11, 0.42{]} \\ \midrule \rowcolor{rowgray}
USA & Yes & {[}-0.14, -0.02{]} & $\emptyset$ &  & {[}$-\infty, +\infty${]} &  & {[}-0.27, 0.12{]} \\
\rowcolor{rowgray} USA & No & $\emptyset$ & $\emptyset$ &
{[}$-\infty, +\infty${]} & {[}$-\infty, +\infty${]} & {[}-0.23,
0.23{]} & {[}-0.22, 0.22{]} \\ \bottomrule
\end{tabular}

\begin{tablenotes}\footnotesize
\item Column "Robust" indicate whether the CS is robust to heteroskedasticity or not. $\emptyset$ indicates an empty CS.
\end{tablenotes}
\end{threeparttable}
}
\end{table}

Several patterns emerge.

First, grid inversion mechanically can enlarge intervals because it includes
entire grid cells containing boundary points rather than precisely locating
the roots of the test statistic. This is visible for the AR test under
heteroskedasticity. For Germany, \citet{Yogo04} reports a robust AR interval
from -1.95 to 1.63, whereas our exact inversion yields -1.73 to 0.66.
Similarly, for Australia under heteroskedasticity, \citet{Yogo04} reports
-0.17 to 0.30, while we obtain -0.11 to 0.22. A comparable phenomenon
appears for the homoskedastic CLR test. For Germany, \citet{Yogo04} reports
-1.23 to 0.28, whereas we obtain -1.18 to 0.24.

Second, grid search may fail to detect sharp or highly localized features of
the confidence interval. Under heteroskedasticity, the AR test for the
United Kingdom illustrates this problem. \citet{Yogo04} reports an interval
from -0.45 to 0.51, whereas our exact inversion yields 0.19 to 0.28. The
grid procedure merges disconnected components and includes values that are
rejected under exact inversion. A similar issue arises for the homoskedastic
LM test in Germany. \citet{Yogo04} reports an interval from -1.21 to 0.26,
while our inversion reveals sharp behavior of the LM statistic near 15.9 and
yields a much wider interval, from -1.18 to 15.91. The coarse grid masks
this nonlinearity.

Third, grid search can misrepresent qualitative properties of the confidence
set, such as emptiness or unboundedness. For the heteroskedastic-robust AR
test in the United States, our exact inversion yields an empty confidence
set, whereas \citet{Yogo04} reports a nonempty interval from -0.14 to -0.02.
In the German homoskedastic LM case discussed above, the grid-based interval
fails to reveal the full extent of the acceptance region. These
discrepancies do not reflect differences in the underlying tests. They arise
solely from numerical inversion.

The Euler equation application is particularly informative because it was
one of the earliest and most influential empirical implementations of grid
inversion for weak identification robust tests. The example therefore shows
that grid-related distortions can arise even in widely studied and carefully
implemented designs.

At the same time, a single application cannot determine whether this
behavior is exceptional or representative. Section 3 revisits grid methods
in a more systematic way, examining 158 empirical specifications spanning
macroeconomics, labor, and public finance to assess how often such failures
occur in practice.

\section{Impact on Applications}

\label{section:applications}

This section evaluates how the Chebyshev approximation performs in applied
settings and compares it to grid-based procedures commonly used in empirical
work. We consider three approaches: (i) the evenly spaced grid search
implemented in the \texttt{weakiv} package, which reflects standard
empirical practice; (ii) a grid based on Chebyshev nodes, which improves
node placement but retains a pure grid-search inversion; and (iii) our
Chebyshev approximation method, which replaces grid search with a global
polynomial approximation.

Whenever exact confidence sets are available, as in the AR, LM, and CQLR
tests, we use them as benchmarks. Our goal is not to reassess the
theoretical properties of these tests, but to study how closely our
approximation reproduces the exact regions and how the two grid-based
methods deviate from them in practice. Coverage comparisons across methods
are not the focus here. Instead, we study numerical accuracy relative to
exact inversion.

For comparability, all grid-based procedures use 501 evaluation points, a
choice that reflects common empirical practice. This ensures that
differences across methods are driven by the inversion approach rather than
by arbitrary choices of grid density. For tests such as CLR and CIL, where
exact inversion is not available and is treated elsewhere in the paper, this
exercise provides reassurance that the approximation behaves as intended,
even though exact confidence sets are not displayed in this section.

We implement the approximation using Chebyshev nodes of the second kind and
a 500-degree polynomial. These nodes are chosen to include the endpoints and
improve approximation accuracy over the entire domain. After compactifying
the parameter space, this ensures that $\beta =0$ and the tails
corresponding to $\beta \rightarrow \pm \infty $ are well represented. This
feature is important because weak identification often produces unbounded
confidence sets, and the behavior of the statistic in the tails determines
whether the region is infinite in each direction.

Maintaining approximate nominal coverage does not require the approximated
confidence set to match the exact analytical form point by point. Different
procedures can achieve the same coverage probability while producing regions
with slightly different shapes. In our empirical designs, such discrepancies
are rare and small, and they do not affect coverage. The approximation
therefore behaves as a controlled perturbation of the exact procedure. For
completeness, we nevertheless quantify how closely the approximation
reproduces the exact geometry whenever a benchmark is available.

To quantify numerical accuracy, we use the Hausdorff distance between
approximate and exact confidence sets. This metric captures the largest gap
between two regions and indicates how much one set must expand to include
the other. A small distance therefore means the approximation closely
matches the exact region, while a large distance signals a meaningful
difference in shape. To make distances comparable across designs and to keep
the scale bounded, all figures report the normalized distance $d/(1+d)$,
which maps $d \in [0,\infty)$ into the unit interval $[0,1]$. Values close
to zero indicate near-perfect agreement with the exact set.

For two sets $A,B\subset \mathbb{R}^{k}$,
\begin{equation}
H(A,B)=\max\left\{ \sup_{x\in A}\inf_{y\in B}\|x-y\|, \sup_{y\in
B}\inf_{x\in A}\|y-x\| \right\}.
\end{equation}
The first term measures the distance from points in $A$ to the closest point
in $B$. The second does the reverse. Taking the maximum makes the measure
symmetric.

We evaluate performance across 158 empirical specifications drawn from five
well-known instrumental variables applications spanning macroeconomics,
labor, and public finance. Three come from the database compiled by %
\citet{AndrewsStockSun19}. \citet{AcconciaCorsettiSimonelli14} study the
effect of public spending on local economic activity, contributing 20
designs. \citet{StephensYang14} estimate returns to schooling using
compulsory schooling laws, contributing 30 designs. \citet{Young14} studies
sectoral employment and productivity using exposure to defense spending,
contributing 40 designs. We also include two classic applications widely
used in the weak-IV literature. \citet{AngristKrueger91} contribute 24
designs based on quarter of birth as an instrument for schooling, and %
\citet{Yogo04} contributes 44 designs estimating the elasticity of
intertemporal substitution across countries and instrument sets.

\begin{figure}[tbph]
\caption{{\protect\small Confidence Set Performance of Chebyshev
Approximation over Grid Algorithms}}
\label{fig:cs_performance}\centering
\begin{minipage}{\textwidth}
        \centering
        \begin{subfigure}[b]{0.49\textwidth}
            \includegraphics[width=\linewidth]{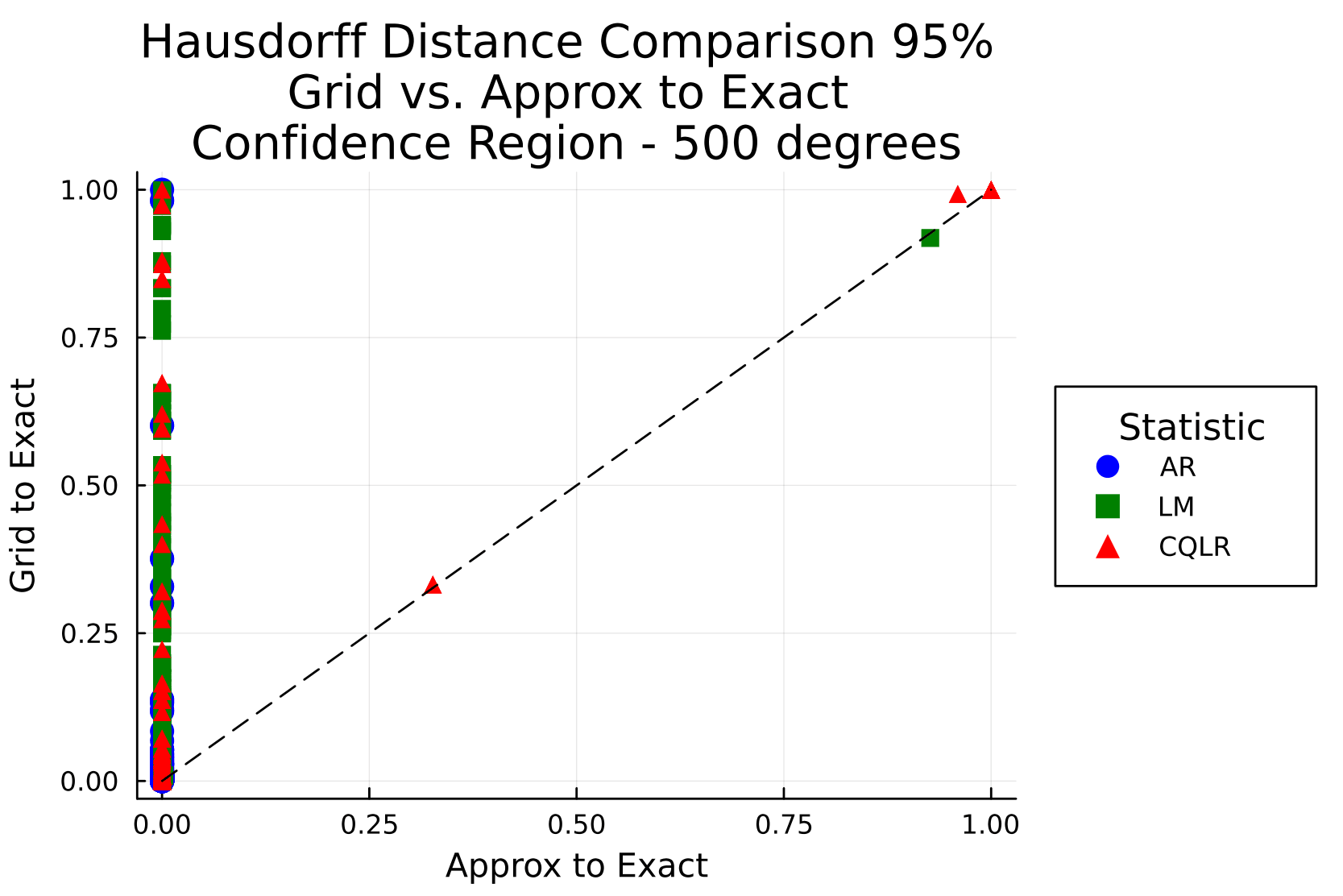}
        \end{subfigure}
        \hfill
        \begin{subfigure}[b]{0.49\textwidth}
            \includegraphics[width=\linewidth]{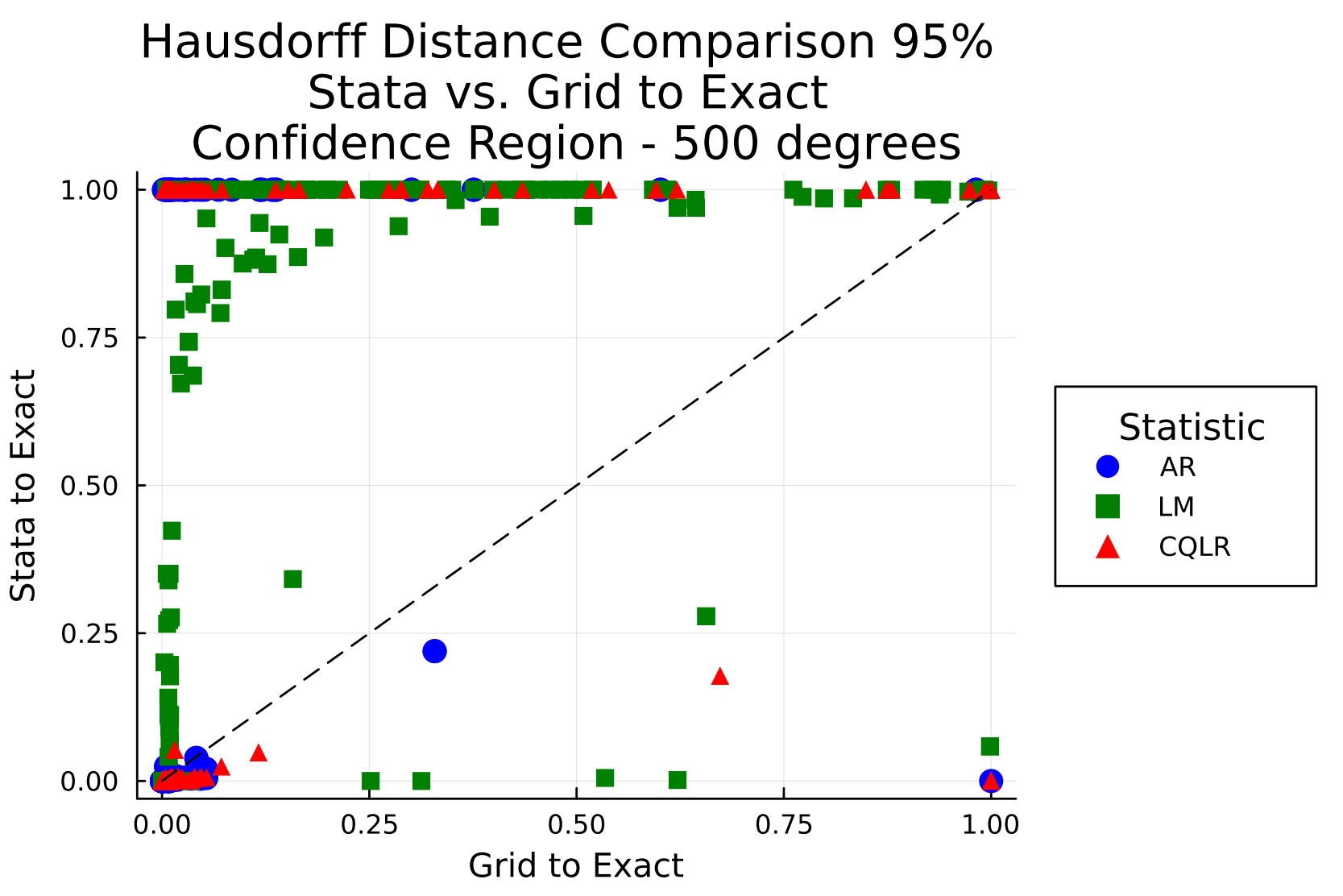}
        \end{subfigure}
\end{minipage}
\end{figure}

Figure~\ref{fig:cs_performance} compares distances design by design using
the full confidence sets. Studying the full sets is important because they
contain more information than their associated confidence intervals, and
differences in shape can reveal numerical errors that are masked once the
region is reduced to an interval. Each point corresponds to one empirical
specification. In the left panel, the horizontal axis reports the normalized
distance $d/(1+d)$ between our Chebyshev approximation and the exact
confidence set, while the vertical axis reports the same distance for a
Chebyshev grid that uses the same nodes but does not apply the polynomial
approximation. Points above the 45-degree line indicate designs for which
the approximation is closer to the exact region. The heavy concentration of
points near zero on the horizontal axis shows that, with only a few rare
exceptions, the approximation is nearly identical to the exact confidence
set. By contrast, the Chebyshev grid alone can exhibit large discrepancies.
The right panel compares grid procedures. The horizontal axis reports the
distance for the Chebyshev grid, while the vertical axis reports the
distance for the evenly-spaced grid implemented in the Stata command \emph{%
weakiv}. Most points above the 45-degree line show that the Chebyshev grid
improves substantially on the standard Stata grid. Although the Chebyshev
grid remains dominated by the full approximation, it performs considerably
better than evenly-spaced grids.

\begin{figure}[tbph]
\caption{{\protect\small Confidence Interval Performance of Chebyshev
Approximation over Grid Algorithms}}
\label{fig:ci_performance}\centering
\begin{minipage}{\textwidth}
        \centering
        \begin{subfigure}[b]{0.49\textwidth}
            \includegraphics[width=\linewidth]{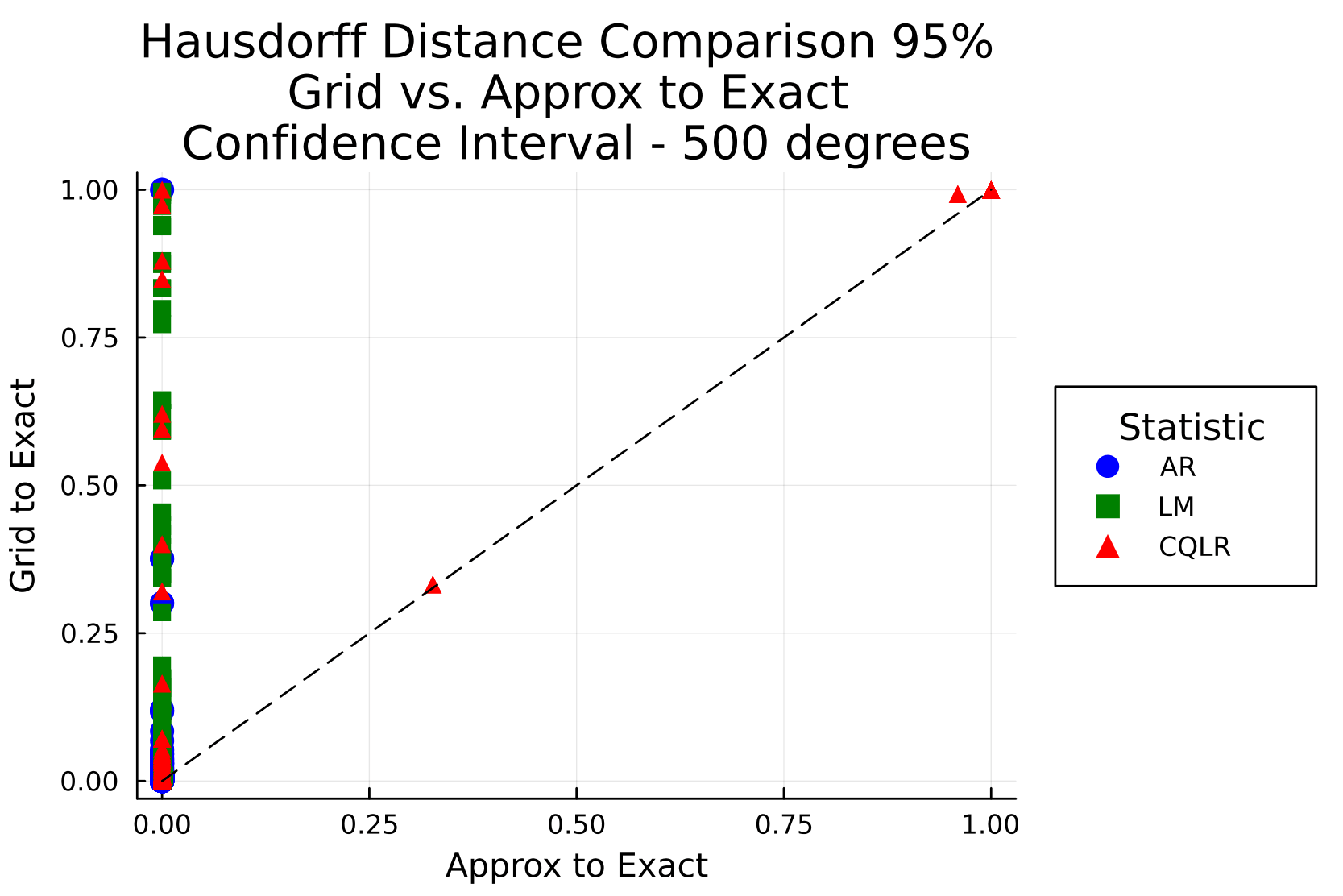}
        \end{subfigure}
        \hfill
        \begin{subfigure}[b]{0.49\textwidth}
            \includegraphics[width=\linewidth]{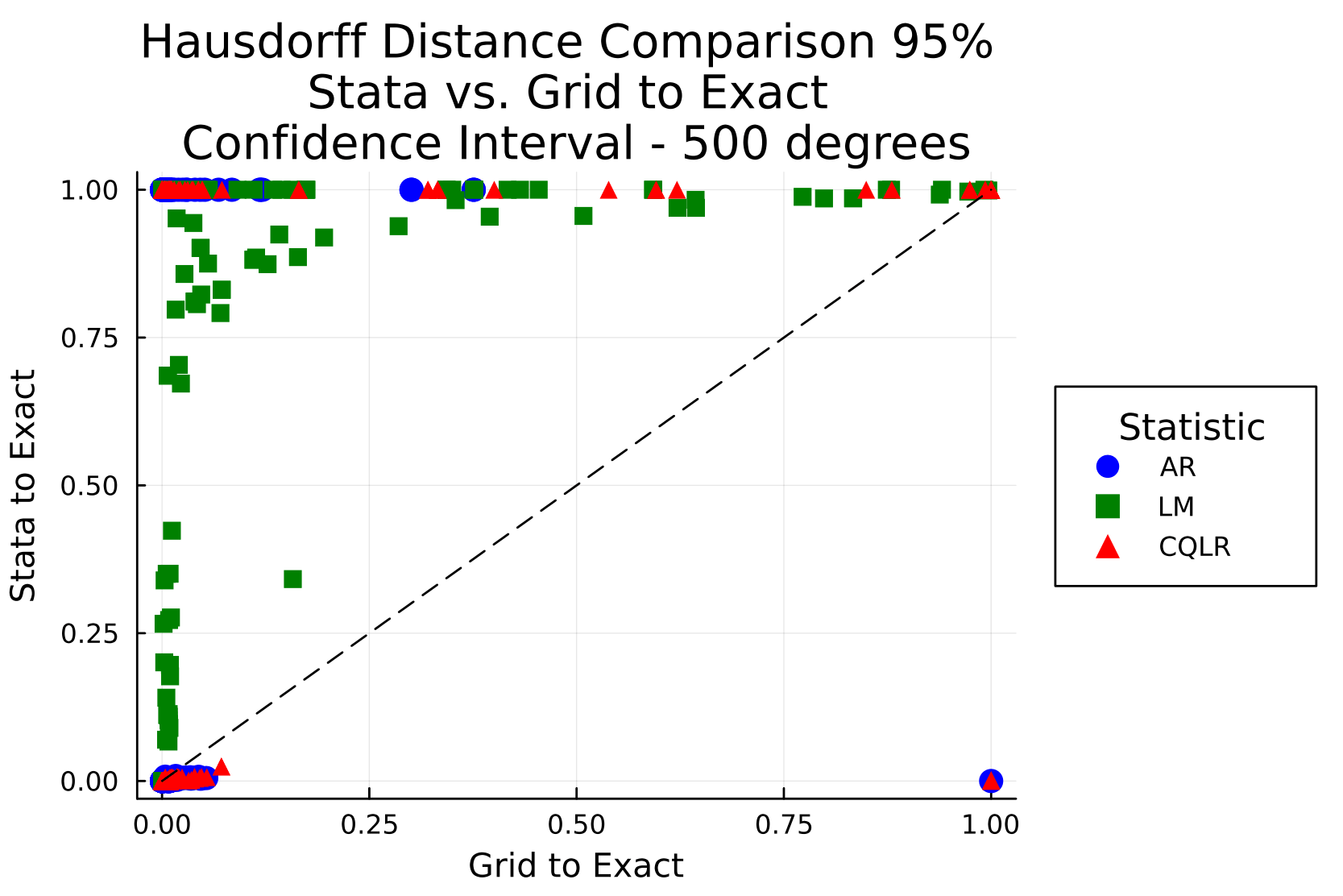}
        \end{subfigure}
\end{minipage}
\end{figure}

Confidence sets robust to weak instruments are not always intervals. The
full region can reveal identification features and show how the data
restrict different parts of the parameter space. In practice, applied
researchers often prefer intervals for communication. We therefore also
examine convex hulls, the smallest intervals containing each confidence set.
Figure~\ref{fig:ci_performance} repeats the comparison for convex hulls,
which correspond to confidence intervals. Confidence intervals are often the
primary object reported in applied work because they summarize uncertainty
in a way that is easy to communicate and compare across specifications. Each
point again represents a design, and the axes have the same interpretation
as in Figure~\ref{fig:cs_performance}. The pattern is similar: the Chebyshev
approximation remains closest to the exact intervals in most designs, while
grid methods frequently produce larger deviations.

\begin{figure}[tbph]
\caption{{\protect\small Stochastic Dominance of Chebyshev Approximation
over Grid Algorithms}}
\label{fig:stochastic_dominance}\centering
\begin{minipage}{\textwidth}
        \centering
        \begin{subfigure}[b]{0.45\textwidth}
            \includegraphics[width=\linewidth]{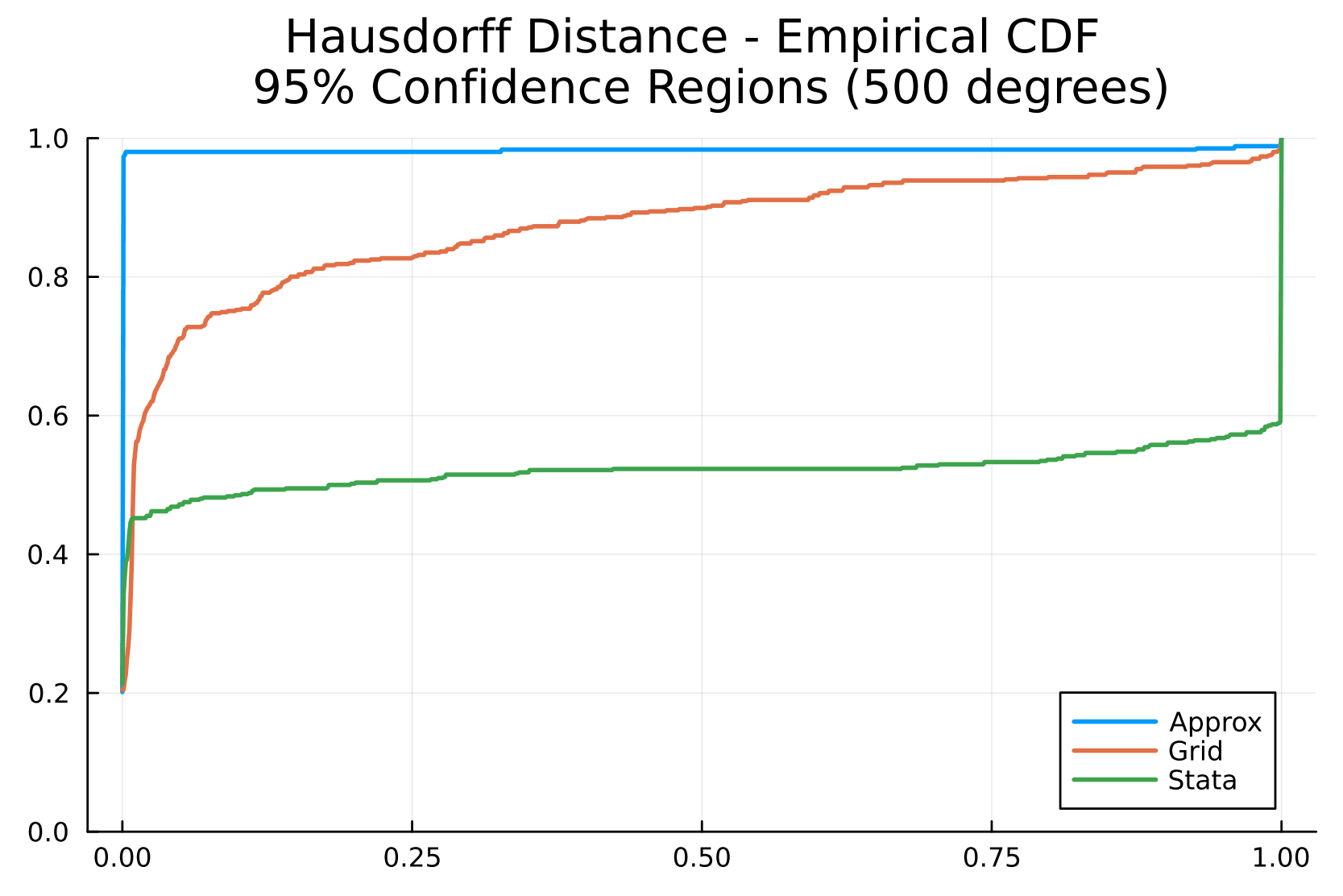}
        \end{subfigure}
        \hfill
        \begin{subfigure}[b]{0.45\textwidth}
            \includegraphics[width=\linewidth]{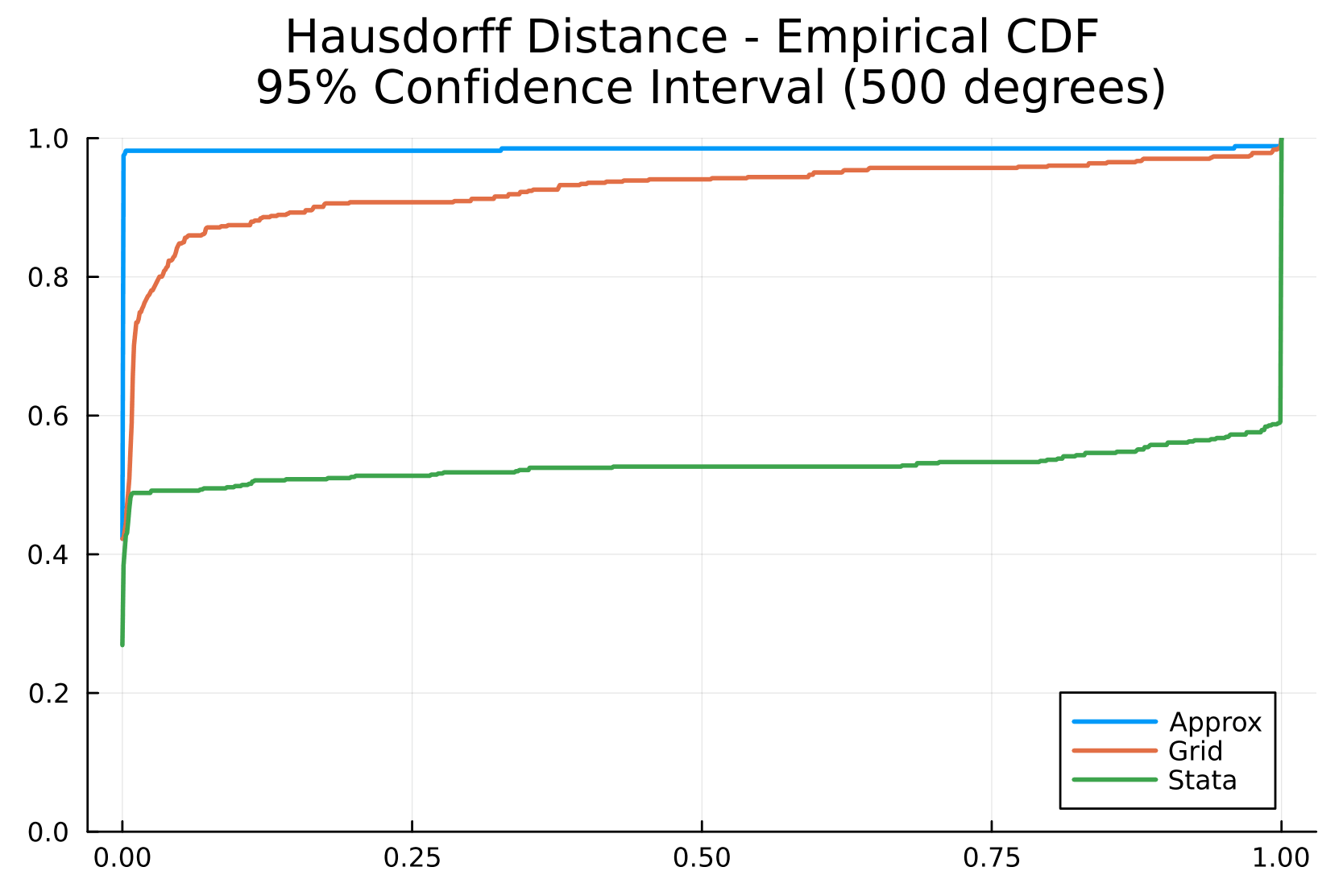}
        \end{subfigure}
\end{minipage}
\end{figure}

Figure~\ref{fig:stochastic_dominance} summarizes performance using empirical
cumulative distribution functions of normalized distances for all three
tests. First-order stochastic dominance shows that the approximation
delivers smaller numerical errors more often. The Chebyshev grid improves on
evenly-spaced grids but remains dominated by the approximation.

The evidence points to clear patterns. The Stata grid method frequently
fails due to how the grid is constructed around the two-stage least squares
(TSLS) estimator, with the search region typically set to $\pm 2$ standard
deviations. When an endpoint of this fixed grid lies inside the confidence
set, the procedure incorrectly extends the set to be unbounded, even when
the true set is bounded. Conversely, when the true confidence set is
unbounded but the grid endpoints fall outside it, the method incorrectly
returns a bounded set. These boundary-driven errors explain why more than
40\% of the specifications in the empirical applications exhibit infinite
Hausdorff distance, corresponding to values equal to one on the x-axis in
Figure~\ref{fig:stochastic_dominance}: the issue is not small numerical
inaccuracy, but a systematic misclassification of whether the confidence set
is informative.

Furthermore, grid methods depend on arbitrary discretization choices and
provide no theoretical bound on approximation error. Replacing evenly-spaced
grids with Chebyshev nodes improves performance, but the choice of nodes
alone does not eliminate large discrepancies. The Chebyshev approximation
adds an additional layer of control and delivers much more reliable recovery
of confidence regions. Whenever exact inversion is available, as in the AR,
LM, and CQLR tests, it should be used in practice. When exact inversion is
not available, as in the CLR and CIL tests, the Chebyshev approximation
provides a reliable and practical alternative that accurately recovers
confidence sets in our empirical applications and avoids the failures of
standard grid methods.

\section{Valid Confidence Regions: Derivation}

\label{section:derivation}

In this section we derive exact confidence sets (CSs) and confidence
intervals (CIs) based on the AR, LM, and CQLR tests. We then construct
approximate CSs and CIs for general conditional tests, including CLR and
CIL. Our goal is to obtain exact confidence regions for the structural
parameter $\beta $. A confidence region is the set of values $\beta _{0}$
for which we do not reject the null hypothesis $H_{0}:\beta =\beta _{0}$
against the two-sided alternative.

These tests depend on the data through the sample second moments. Recall
from Equation (\ref{eq:model}) that $y_{1i}$ denotes the outcome variable, $%
y_{2i}$ the endogenous regressor of interest, $z_i$ the $k\times 1$ vector
of instruments, and $x_i$ the $d\times 1$ vector of included exogenous
covariates. Inference is based on low-dimensional functions of these
observable variables.

\begin{equation}
\frac{1}{n}\sum_{i=1}^{n} z_i z_i^{\prime}, \quad \frac{1}{n}\sum_{i=1}^{n}
z_i x_i^{\prime}, \quad \frac{1}{n}\sum_{i=1}^{n} x_i x_i^{\prime},
\label{eq:LLNs}
\end{equation}
as well as the scaled sample moments
\begin{equation}
\frac{1}{\sqrt{n}}\sum_{i=1}^{n} z_i y_{1i}, \quad \frac{1}{\sqrt{n}}%
\sum_{i=1}^{n} z_i y_{2i}, \quad \frac{1}{\sqrt{n}}\sum_{i=1}^{n} z_i
x_i^{\prime}.  \label{eq:CLTs}
\end{equation}

The moments in (\ref{eq:LLNs}) and (\ref{eq:CLTs}) arise directly from the
IV orthogonality condition. Under the null hypothesis $\beta=\beta_0$, the
structural residual $y_{1i}-\beta_0 y_{2i}-x_i^{\prime }\gamma_1$ must be
uncorrelated with the instruments. The scaled sample moments in (\ref%
{eq:CLTs}) therefore collect the empirical covariances between the
instruments and the outcome, the endogenous regressor, and the included
controls. The second moments in (\ref{eq:LLNs}) provide the normalization
and covariance structure required to partial out the controls and to form
quadratic test statistics. Together, these objects contain all information
in the sample that is relevant for inference on $\beta$.

Inference relies on laws of large numbers for the averages in (\ref{eq:LLNs}%
) and central limit theorems for the statistics in (\ref{eq:CLTs}), allowing
for heteroskedasticity, autocorrelation, or general dependence. Under
standard regularity conditions, the statistics in (\ref{eq:CLTs}), properly
centered and scaled, are asymptotically normal with well-defined covariance
matrices. We take these asymptotic results as given and focus on expressing
the tests in terms of a lower-dimensional sufficient statistic.

The tests depend on (\ref{eq:LLNs}) and (\ref{eq:CLTs}) only through a much
smaller set of transformed statistics. This reduction removes the nuisance
coefficients on the covariates by projecting the instruments onto the space
orthogonal to the exogenous regressors and standardizing the result. After
these two steps, inference depends only on a low-dimensional statistic.

To describe the reduction precisely, consider the linear IV model (\ref%
{eq:model}) written in matrix form
\begin{eqnarray*}
y_{1} &=& y_{2}\beta + X\gamma_{1} + u, \\
y_{2} &=& Z\pi + X\gamma_{2} + v_{2},
\end{eqnarray*}
where $y_{1}$ and $y_{2}$ are $n\times 1$ vectors, $Z$ and $X$ are $n\times
k $ and $n\times d$ matrices of instruments and exogenous covariates with
full column rank, and $u$ and $v_{2}$ are zero mean errors. The reduced form
for $Y=[y_{1},y_{2}]$ can be written as
\begin{equation*}
Y = Z\pi a^{\prime} + X\Gamma + V,
\end{equation*}
where $a=(\beta,1)^{\prime}$, $\Gamma=[\gamma_{1},\gamma_{2}]$, and $%
V=[v_{1},v_{2}]=[u+\beta v_{2},v_{2}]$.

We partial out the covariates by regressing the instruments on $X$ and
working with the residuals. Let
\begin{equation*}
Z_{\perp }=M_{X}Z,\quad M_{X}=I-N_{X},\quad N_{X}=X(X^{\prime
}X)^{-1}X^{\prime },
\end{equation*}%
and write $z_{i}^{\perp \prime }$ for the $i$th row of $Z_{\perp }$. The
relevant orthonormalized moments are
\begin{equation*}
\left( \sum_{i=1}^{n}z_{i}^{\perp }z_{i}^{\perp \prime }\right)
^{-1/2}\sum_{i=1}^{n}z_{i}^{\perp }y_{1i},\quad \left(
\sum_{i=1}^{n}z_{i}^{\perp }z_{i}^{\perp \prime }\right)
^{-1/2}\sum_{i=1}^{n}z_{i}^{\perp }y_{2i}.
\end{equation*}

These moments can be written compactly as
\begin{equation}
R = (Z_{\perp}^{\prime}Z_{\perp})^{-1/2} Z_{\perp}^{\prime} Y.  \label{eq:R}
\end{equation}

The statistic $R$ summarizes all information in (\ref{eq:LLNs}) and (\ref%
{eq:CLTs}) that is relevant for inference on $\beta$. This transformation
eliminates dependence on nuisance coefficients and yields a low-dimensional
sufficient statistic. Under standard conditions, a central limit theorem
implies that $R$ is asymptotically normal with covariance matrix $\Sigma$. A
consistent estimator $\widehat{\Sigma}_{n}$ can be constructed using
standard variance estimators. See \citet{White80} for heteroskedastic data, %
\citet{NeweyWest87} and \citet{Andrews91} for heteroskedastic and
autocorrelated data, and \citet{CameronGelbachMiller11} for clustered data.
A general textbook overview is given by \citet{Hansen22}.

We study confidence sets based on tests that are robust to weak
identification: Anderson Rubin (AR), Lagrange multiplier (LM), conditional
quasi likelihood ratio (CQLR), conditional likelihood ratio (CLR), and
conditional integrated likelihood (CIL). We first develop exact numerical
methods for AR, LM, and CQLR. These tests control size at level $\alpha$
under weak identification and therefore generate confidence sets with exact
coverage. The resulting confidence sets can be unbounded with positive
probability, an important empirical feature of weak identification. From now
on, $1-\alpha$ denotes the target coverage probability, typically 95
percent. We then propose an approximation method for more general
conditional tests, including CLR and CIL, that achieves coverage arbitrarily
close to $1-\alpha$.

\subsection{The AR Confidence Region}

The Anderson-Rubin (AR) test, introduced by \citet{AndersonRubin49} under
homoskedasticity and extended to general GMM settings by %
\citet{StockWright00}, is based on the moment condition $\mathbb{E}%
[Z^{\prime }u]=0$ evaluated under the null hypothesis. Under the null
hypothesis, the AR statistic has an asymptotic $\chi _{k}^{2}$ pivotal
distribution regardless of the strength of the instruments.

For a candidate value $\beta _{0}$, the AR statistic is
\begin{equation}
AR(\beta _{0})=b_{0}^{\prime }R^{\prime }\left[ (b_{0}^{\prime }\otimes
I_{k})\widehat{\Sigma }_{n}(b_{0}\otimes I_{k})\right] ^{-1}Rb_{0},
\label{eq:AR}
\end{equation}%
where $b_{0}=(1,-\beta _{0})^{\prime }$. This quadratic form corresponds to
the sample moment
\begin{equation}
\frac{1}{n}\sum_{i=1}^{n}z_{i}^{\perp }(y_{1i}-\beta _{0}y_{2i}),
\label{eq:ARmoment}
\end{equation}%
that is, the structural residual evaluated at $\beta _{0}$. The covariance
matrix in (\ref{eq:AR}) is estimated at the same parameter value at which
the moment in (\ref{eq:ARmoment}) is evaluated, so the weighting matrix
depends on $\beta _{0}$. The criterion is therefore continuously updating.

\citet{MoreiraNeweySharifvaghefi24} apply the Sherman-Morrison formula to
show that the matrix
\begin{equation*}
\left[ (b_{0}^{\prime }\otimes I_{k})\widehat{\Sigma }_{n}(b_{0}\otimes
I_{k})\right] ^{-1}
\end{equation*}%
is a ratio of matrix valued polynomials in $\beta _{0}$. The numerator is a
degree $(2k-2)$ matrix polynomial and the denominator is a degree $2k$
scalar polynomial. Because $Rb_{0}$ is linear in $\beta _{0}$, the AR
statistic itself can be written as a ratio of degree $2k$ polynomials,
\begin{equation*}
AR(\beta _{0})=\frac{p_{AR}(\beta _{0})}{q_{AR}(\beta _{0})}=\frac{\gamma
_{0}+\gamma _{1}\beta _{0}+\cdots +\gamma _{2k}\beta _{0}^{2k}}{\delta
_{0}+\delta _{1}\beta _{0}+\cdots +\delta _{2k}\beta _{0}^{2k}}.
\end{equation*}

Therefore inversion of the AR test reduces to solving a polynomial
inequality. This algebraic structure is the key computational
simplification. Without loss of generality we normalize $\delta _{0}=1$. The
coefficients of these polynomials are obtained numerically by evaluating the
statistic at sufficiently many values of $\beta _{0}$ and solving a linear
system.

Let $c_{\alpha }(k)$ denote the $(1-\alpha )$ quantile of the $\chi _{k}^{2}$
distribution. The AR confidence set is
\begin{equation*}
CS_{AR}=\left\{ \beta _{0}:\frac{p_{AR}(\beta _{0})}{q_{AR}(\beta _{0})}\leq
c_{\alpha }(k)\right\} .
\end{equation*}%
The boundary of the confidence set is obtained by solving the degree 2k
polynomial equation%
\begin{equation*}
p_{AR}(\beta _{0})-c_{\alpha }(k)q_{AR}(\beta _{0})=0.
\end{equation*}%
All roots can be computed using standard numerical routines. Figure \ref%
{fig:AR_graph} illustrates a typical realization of the AR statistic for $%
k=10$. After computing all boundary points, we evaluate the statistic at
midpoints of the induced intervals to determine exactly which intervals
belong to the confidence set.

\begin{figure}[h!]
\caption{Confidence sets for AR test}
\label{fig:AR_graph}\centering
\vspace{10pt} \includegraphics[width=0.8\linewidth]{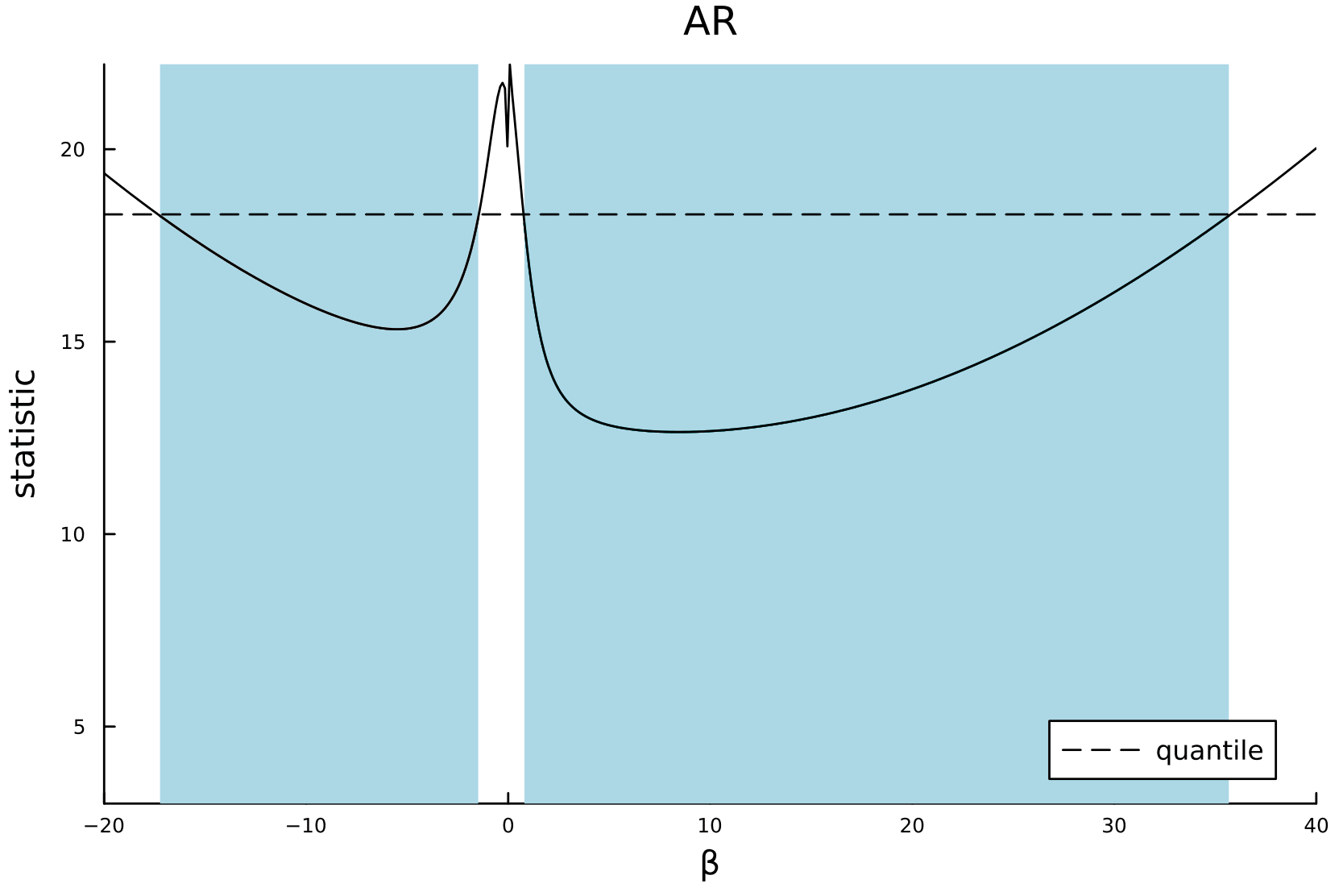}
\end{figure}

While Figure \ref{fig:AR_graph} appears to show only two confidence
intervals, the test statistic actually fluctuates sharply in the middle.
Because our method is exact, it identifies a third, very small interval that
would typically be missed by a standard grid search or simple visual
inspection. Figure \ref{fig:AR_graph} zooms in on this area to reveal this
hidden component, with all numerical bounds detailed in Table \ref%
{table:AR_CS}.

\begin{table}[h]
\caption{Interval components of the AR CS in Figure~\protect\ref%
{fig:AR_graph}}
\label{table:AR_CS}\centering \vspace{10pt}
\begin{tabular}{cccc}
\toprule & Interval 1 & Interval 2 & Interval 3 \\
\midrule Lower Bound & -17.333 & -0.020 & 0.758 \\
Upper Bound & -1.454 & 0.002 & 35.758 \\
\bottomrule &  &  &
\end{tabular}%
\end{table}

\begin{figure}[h!]
\caption{Detail of Figure \protect\ref{fig:AR_graph}}
\label{fig:AR_zoom}\centering
\vspace{10pt} \includegraphics[width=0.8\linewidth]{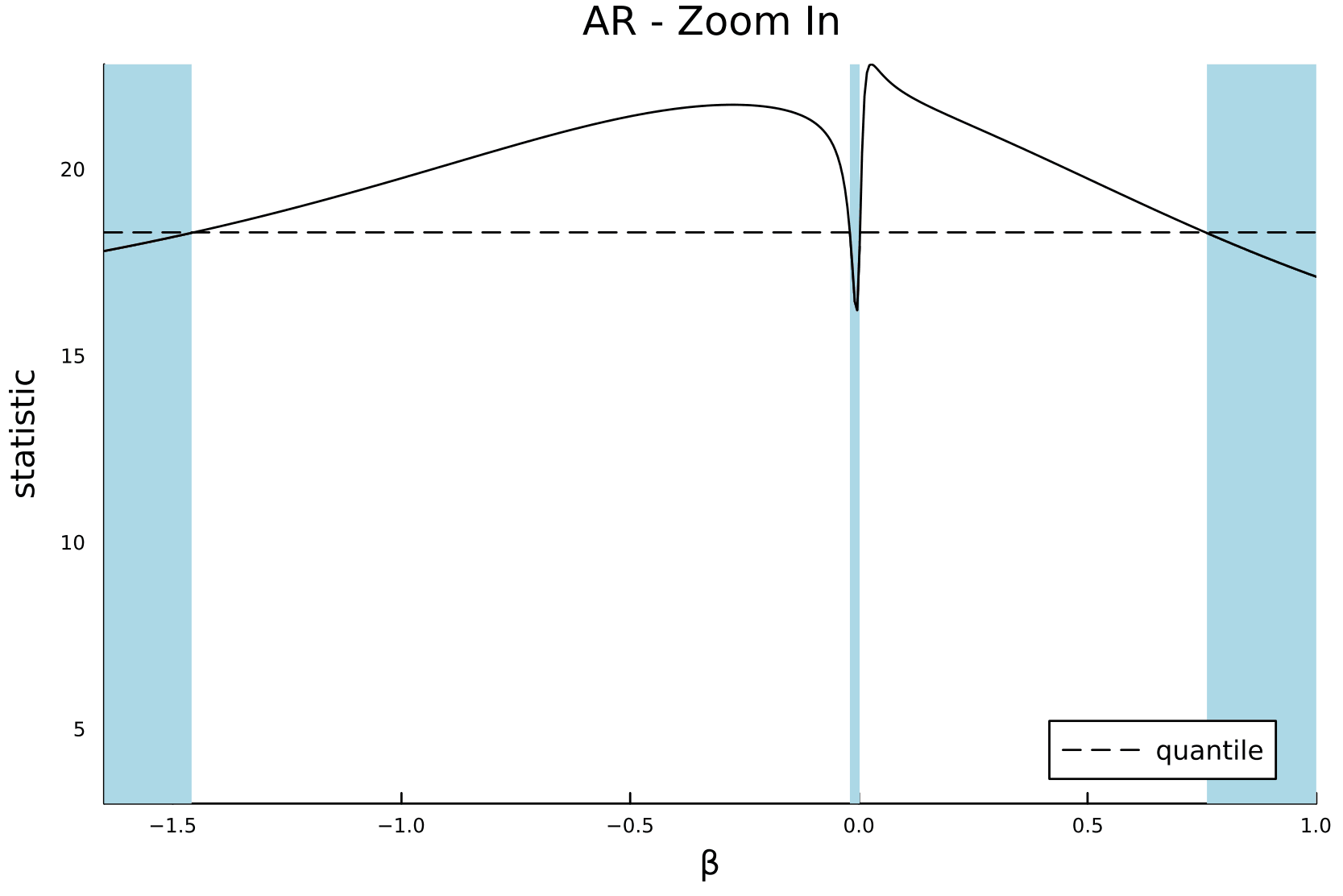}
\end{figure}

\subsection{The LM Confidence Region}

The LM statistic can be interpreted as a score test. Intuitively, it
measures how sensitive the likelihood of the statistic $R$ is to small
deviations from the null value $\beta_0$. The statistic is obtained from the
derivative of the Gaussian log likelihood of $R$ with respect to $\beta$,
evaluated at $\beta_0$, and normalized by an estimate of its variance.

\citet{AndrewsMoreiraStock04} and \citet{Kleibergen05} show that the LM
statistic asymptotically has a pivotal $\chi _{1}^{2}$ distribution under
the null hypothesis, even under weak identification. Inverting the LM test
therefore yields valid confidence sets based on the same sufficient
statistic $R$ used in the AR case.

Formally, the LM statistic at a candidate value $\beta _{0}$ is
\begin{equation}
LM(\beta _{0})=\frac{p_{LM}(\beta _{0})^{2}}{q_{LM}(\beta _{0})},
\label{eq:LM}
\end{equation}%
where the numerator
\begin{equation*}
p_{LM}(\beta _{0})=\text{vec}(R)^{\prime }(b_{0}\otimes
I_{k})[(b_{0}^{\prime }\otimes I_{k})\widehat{\Sigma }_{n}(b_{0}\otimes
I_{k})]^{-1}[(a_{0}^{\prime }\otimes I_{k})\widehat{\Sigma }%
_{n}{}^{-1}(a_{0}\otimes I_{k})]^{-1}(a_{0}^{\prime }\otimes I_{k})\widehat{%
\Sigma }_{n}{}^{-1}\text{vec}(R)
\end{equation*}%
is the profile score of the Gaussian log likelihood for $R$ evaluated at $%
\beta _{0}$, and the denominator
\begin{eqnarray*}
q_{LM}(\beta _{0}) &=&\text{vec}(R)^{\prime }(a_{0}\otimes
I_{k})[(a_{0}^{\prime }\otimes I_{k})\widehat{\Sigma }_{n}{}^{-1}(a_{0}%
\otimes I_{k})]^{-1}[(b_{0}^{\prime }\otimes I_{k})\widehat{\Sigma }%
_{n}(b_{0}\otimes I_{k})]^{-1} \\
&&\times \lbrack (a_{0}^{\prime }\otimes I_{k})\widehat{\Sigma }%
_{n}{}^{-1}(a_{0}\otimes I_{k})]^{-1}(a_{0}^{\prime }\otimes I_{k})\widehat{%
\Sigma }_{n}{}^{-1}\text{vec}(R)
\end{eqnarray*}%
is an estimator of the asymptotic variance of that score.

Although the expression is algebraically more involved than in the AR case,
the computational structure is the same. The key computational point is
that, like the AR statistic, the LM statistic is a rational function of $%
\beta _{0}$. The Sherman Morrison formula implies that all matrix terms
above can be written as ratios of matrix polynomials in $\beta _{0}$, and
all linear forms in $\text{vec}(R)$ depend linearly on $\beta _{0}$. As a
result, the LM statistic itself can be written as a ratio of finite degree
polynomials in $\beta _{0}$. The maximal degree is $8k-4$. Finite degree
implies a finite number of real boundary points, so the confidence set can
be recovered exactly by enumerating all roots.

Exactly as in the AR case, this algebraic structure reduces inversion to
solving a polynomial inequality in $\beta_0$. The coefficients of these
polynomials are obtained numerically by evaluating the statistic at
sufficiently many values of $\beta_0$ and solving a linear system. No
symbolic manipulation is required.

Let $c_\alpha(1)$ denote the $(1-\alpha)$ quantile of the $\chi^2_1$
distribution. The LM confidence set is
\begin{equation*}
CS_{LM} = \left\{ \beta_0 : \frac{p_{LM}(\beta_0)^2}{q_{LM}(\beta_0)} \le
c_\alpha(1) \right\}.
\end{equation*}

The boundary of this set is obtained by solving the polynomial equation
\begin{equation*}
p_{LM}(\beta_0)^2 - c_\alpha(1) q_{LM}(\beta_0) = 0.
\end{equation*}
All real roots of this polynomial are computed numerically. We then evaluate
the statistic at midpoints between consecutive roots to determine which
intervals belong to the confidence set.

\begin{figure}[h]
\caption{Confidence sets for LM test}
\label{fig:LM_graph}\centering
\vspace{10pt} \includegraphics[width=0.8\linewidth]{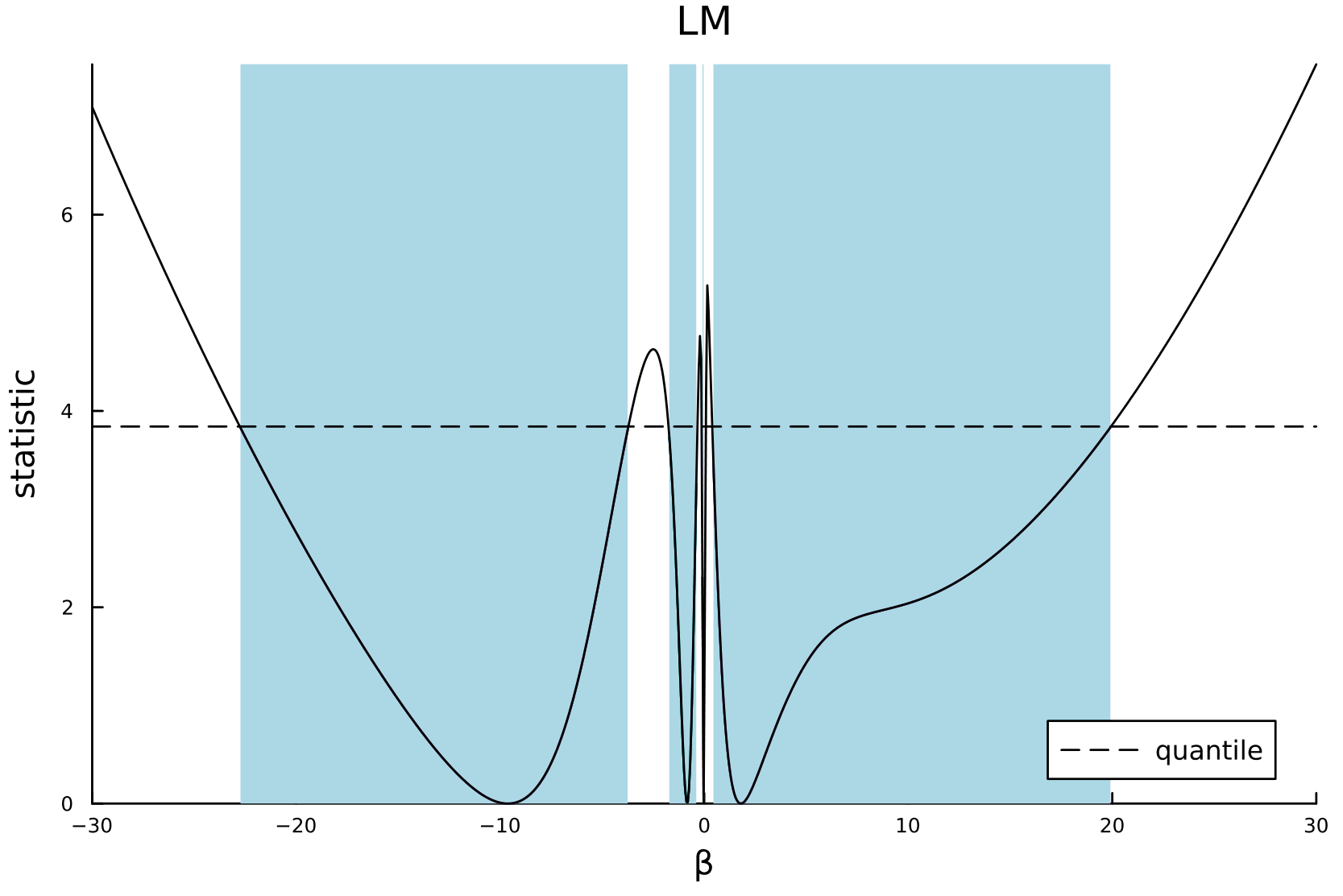}
\end{figure}

Figure \ref{fig:LM_graph} illustrates a typical realization for $k=10$. The
LM statistic can exhibit highly non-monotonic behavior in parts of the
parameter space, generating many disjoint components. Table \ref{table:LM_CS}
shows that the confidence set contains arbitrarily small intervals near zero
and also a large interval far from the true value $\beta =0$. Any grid based
inversion with a finite resolution would necessarily miss some of these
components, especially the arbitrarily small ones. Our method recovers all
of them by construction.

\begin{table}[h]
\caption{Interval components of the LM CS in Figure~\protect\ref%
{fig:LM_graph}}
\label{table:LM_CS}\centering \vspace{10pt}
\begin{tabular}{ccccccc}
\toprule & Interval 1 & Interval 2 & Interval 3 & Interval 4 & Interval 5 &
Interval 6 \\
\midrule Lower Bound & -23180.152 & -22.763 & -1.751 & -0.122 & 0.008 & 0.392
\\
Upper Bound & -18742.755 & -3.721 & -0.334 & 0.003 & 0.083 & 19.935 \\
\bottomrule &  &  &  &  &  &
\end{tabular}%
\end{table}

\subsection{CQLR Confidence Region}

\label{section:CQLR}

\citet{AndrewsMoreiraStock04} and \citet{Kleibergen05} adapt the likelihood
ratio statistic from the homoskedastic linear IV model to settings with HAC
errors and to the general GMM framework, respectively. The QLR statistic
specialized for the HAC-IV model becomes
\begin{equation}
QLR(\beta _{0})=\frac{1}{2}\left[ AR(\beta _{0})-r(\beta _{0})+\sqrt{\left[
AR(\beta _{0})-r(\beta _{0})\right] ^{2}+4\,LM(\beta _{0})\cdot r(\beta _{0})%
}\right] ,  \label{eq:QLR}
\end{equation}%
where $AR(\beta _{0})$ is defined in \eqref{eq:AR}, $LM(\beta _{0})$ is
defined in \eqref{eq:LM}, and $r(\beta _{0})$ is a rank statistic.

Rank statistics depend on a weighted orthogonalization of the sample
Jacobian of the moment conditions in such a way that the orthogonalization
creates a statistic asymptotically independent of the sample moments (see
\cite{AndrewsGuggenberger17}). Intuitively, the rank statistic isolates the
information provided by the instruments about the endogenous regressor,
netting out the effect of the structural parameter being tested. These rank
statistics are suitable for testing rank conditions. In the linear IV
setting, the rank restriction can be written as $\text{rank}\,\mathbb{E}%
(z_{i}y_{2i})=0$, which is equivalent to testing $\pi =0$. Different rank
statistics generate different CQLR tests.

One particular choice of rank statistic is
\begin{equation}
r(\beta _{0})=\text{vec}(R)^{\prime }\widehat{\Sigma }_{n}{}^{-1}(a_{0}%
\otimes I_{k})\left[ (a_{0}^{\prime }\otimes I_{k})\widehat{\Sigma }%
_{n}{}^{-1}(a_{0}\otimes I_{k})\right] ^{-1}(a_{0}^{\prime }\otimes I_{k})%
\widehat{\Sigma }_{n}{}^{-1}\text{vec}(R),  \label{eq:rank}
\end{equation}%
where $a_{0}=(\beta _{0},1)^{\prime }$. Under Gaussian $R$, this rank
statistic is related to an estimator of $\pi $ under the null $\beta =\beta
_{0}$. This rank statistic has an important property on which we rely to
develop our method: it is a ratio of polynomials in $\beta _{0}$. This
algebraic structure is central because it allows us to map sets in the rank
domain back into the parameter domain by solving polynomial equations. We
proceed using \eqref{eq:rank}, but the method generalizes to any rational
rank statistic.

Unlike the AR and LM statistics, the likelihood ratio (LR), quasi-likelihood
ratio (QLR), and integrated likelihood (IL) statistics are not pivotal, so %
\citet{Moreira03} proposes replacing the fixed chi-square critical value by
a conditional critical value. For the CQLR test, \citet{Moreira03} and %
\citet{Kleibergen05} show that the conditional critical values depend on the
data only through the rank statistic. If we denote by $\kappa _{\alpha
}(r(\beta _{0}))$ the critical value function (CVF) when we observe $r(\beta
_{0})$, the confidence set is given by the solution of
\begin{equation}
QLR(\beta _{0})\leq \kappa _{\alpha }(r(\beta _{0})).  \label{eq:CQLR_CS}
\end{equation}

In this case, we cannot solve for the boundary points of the confidence set
by solving a single polynomial equation. Because the critical value varies
with the data, we cannot use a flat horizontal threshold as in the pivotal
chi-square case. The difficulty in inverting conditional tests is that their
CVFs do not admit a simple algebraic representation in $\beta_0$.

To motivate our algorithm, it is useful to recall the structure under
homoskedasticity. In this case, $\widehat{\Sigma}_n$ has a particularly
simple form: it is built from a $2\times 2$ matrix $\widehat{\Omega}_n$ that
captures the covariance of the reduced-form errors, repeated in the same way
across all instruments. The QLR statistic simplifies to a function of the
rank statistic,
\begin{equation*}
QLR(\beta _{0})=\lambda _{\max }-r(\beta _{0}),
\end{equation*}%
where $\lambda _{\max }$ is the largest eigenvalue of $\widehat{\Omega }%
_{n}^{-1/2}R^{\prime }R\widehat{\Omega }_{n}^{-1/2}$ and depends only on the
data, while $r(\beta _{0})$ is a ratio of quadratic polynomials in $\beta
_{0}$.

\begin{figure}[tbp]
\caption{First and second derivative of the CVF, as function of the rank
statistic}
\label{fig:CVF_convexity}\centering \vspace{5mm} \includegraphics[width=1%
\linewidth]{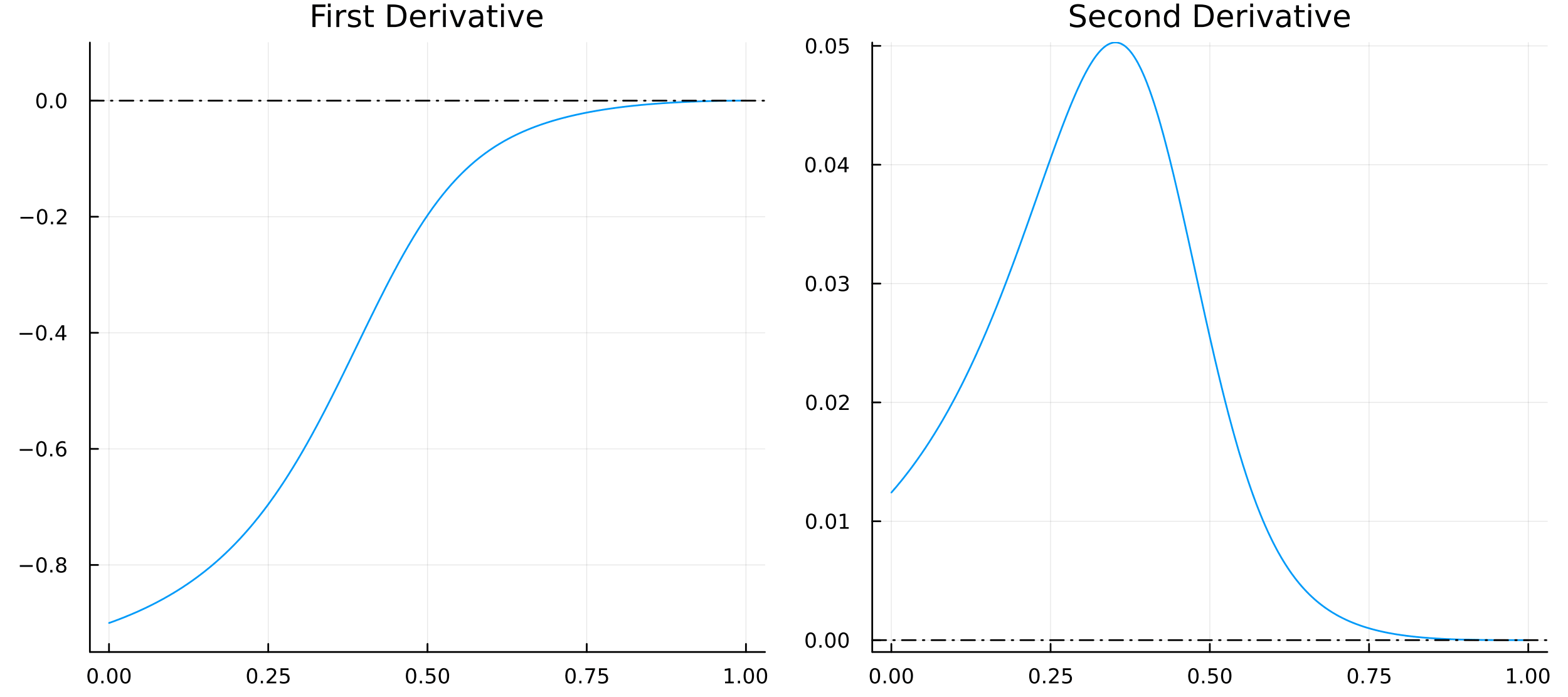}
\par
{\footnotesize Note: the domain $[0,+\infty)$ of the rank statistic is
compactified on $[0,1]$ via the transformation $r \mapsto (2/\pi)\cdot
\tan^{-1}(r/20)$. }
\end{figure}

\citet{Mikusheva10} exploits this simplification to construct a threshold
method that exactly recovers the confidence set. The key idea is to view the
inequality defining the confidence set as an inequality between two
functions of the rank statistic, rather than as a direct inequality in $%
\beta_0$. The reason is that the behavior of the CVF as a function of $%
\beta_0$ is complicated, while both the statistic and the CVF are well
behaved as functions of $r$: $QLR(r)$ is linear, and $\kappa_\alpha(r)$ is
strictly decreasing (\citet{Moreira03} and \citet{Mikusheva10}) and strictly
convex (Figure~\ref{fig:CVF_convexity} illustrates these properties for $%
k=10 $, and Appendix Section~\ref{supp:CQLR} extends the analysis to other
values of $k$.). The algorithm becomes: (i) solve for the values of $r$
satisfying $\lambda_{\max}-r\leq \kappa_\alpha(r)$; (ii) recover the values
of $\beta_0$ using the rational structure of $r(\beta_0)$. Under
homoskedasticity, the function $\kappa_\alpha(r)+r$ is strictly increasing.
Therefore, the confidence set can be characterized by $r(\beta_0)\geq r^*$,
where $r^*$ solves $r^*+\kappa_\alpha(r^*)=\lambda_{\max}$. The threshold $%
r^*$ (if it exists) can be found by bisection, so inversion reduces to a
simple intersection problem.\footnote{%
Practitioners need to be careful, because the value $r^*$ does not always
exist. \citet{Moreira03} shows that the CVF is bounded above by $%
q_{\alpha}(k)$, which implies that a threshold $r^*$ exists if and only if $%
\lambda_{\max}\geq q_{\alpha}(k)$. Otherwise, the confidence set is the
whole real line. Other possible shapes for the confidence set are $%
(-\infty,x_{1}]\cup[x_{2},+\infty)$ and $[x_{1},x_{2}]$.}

This approach cannot be directly generalized to HAC settings. Under HAC
errors, the QLR statistic does not admit a global linear representation in $%
r(\beta_0)$, and the geometric symmetry of the homoskedastic case breaks
down. We overcome this difficulty with a piecewise approach. We split the
parameter space into intervals on which $r(\beta_0)$ is a bijection. Within
each such interval, the QLR statistic becomes an implicit function of the
rank statistic. We then solve for the values of $r$ satisfying the
confidence set inequality $QLR(r)\leq \kappa_\alpha(r)$, obtaining a finite
union of intervals for $r$. Finally, for each valid interval for $r$, we map
back to $\beta_0$ using the rational structure of $r(\beta_0)$.

A remaining difficulty is that solving $QLR(r)\leq \kappa_\alpha(r)$ for $r$
can be nontrivial when errors are not homoskedastic. We can find all roots
of the difference between two functions if we know the points at which they
change their monotonicity or curvature. Mapping out these shape changes
allows us to bound the functions and locate all intersections without
missing hidden dips or disconnected components. This is another reason for
working in the rank domain: although we do not know much about the
composition $\kappa_\alpha(r(\beta))$, we do know that $\kappa_\alpha(r)$ is
strictly decreasing and strictly convex. The homoskedastic case can be
viewed as a highly symmetric benchmark in which the QLR is globally linear
in the rank statistic. Our method extends the reliability of that case to
HAC settings by handling the nonlinearities piecewise.

To implement exact inversion under HAC errors, we must address two
difficulties: (i) the mapping $r(\beta_0)$ is only piecewise invertible in $%
\beta_0$, and (ii) the QLR statistic may change monotonicity or curvature
within each piece. Exact inversion therefore proceeds by first decomposing
the parameter space into regions where the geometry is well-behaved and then
locating all intersections in the rank domain before mapping back to $%
\beta_0 $. Formally, the procedure consists of four steps:

\begin{enumerate}
\item Partition the parameter space into maximal intervals on which $r(\beta
)$ is injective. On each such interval the mapping between $\beta $ and $r$
is one-to-one, allowing us to treat boundary points as geometric
intersections in the rank domain (see Section~\ref%
{section:injective_monotone_convex}).

\item Within each interval from Step 1, determine all points at which the
QLR statistic changes monotonicity or curvature as a function of the rank
statistic. This maps out the exact shape of the test statistic in the rank
domain (see Section~\ref{section:injective_monotone_convex}).

\item For each interval generated by the previous steps, apply the root
finding procedure described in Section~8.2 to compute all solutions of $%
QLR(r)\leq \kappa _{\alpha }(r)$ inside that interval.

\item After collecting the valid intervals for the rank statistic, map them
back to the structural parameter $\beta _{0}$ using the rational
representation of $r(\beta _{0})$.
\end{enumerate}

The appendix provides a complete algorithmic implementation of these steps,
including proofs that all boundary points are recovered and no components of
the confidence set are missed.

\subsection{General Conditional Tests}

\label{section:general_big_picture}

We now extend our analysis to general conditional tests under HAC errors.
Examples include the Conditional Likelihood Ratio (CLR) test and the
Conditional Integrated Likelihood (CIL) test. These procedures often deliver
substantial power gains relative to AR in over-identified settings, but they
are considerably more difficult to invert. Our goal in this section is to
provide a general, numerically stable method for constructing reliable
confidence sets based on such tests.

\paragraph{The CLR Test.}

The likelihood ratio statistic underlying the CLR test is
\begin{equation}
LR(\beta_0)=\sup_{\beta} r(\beta)-r(\beta_0),  \label{eq:LR_statistic}
\end{equation}
where $r(\beta)$ is the rank statistic defined in (\ref{eq:rank}).

Under homoskedasticity, the CQLR and the CLR are numerically identical.
Under HAC errors, however, the CLR allows for a more flexible treatment of
the covariance structure. \citet{AndrewsMikusheva16} develop the CLR test
for general nonlinear GMM settings. In the linear IV model, %
\citet{MoreiraMoreira19} analyze the CLR but do not provide a complete
computational implementation, as the statistic requires solving a supremum
over the parameter space at each evaluation point. This nested optimization
makes direct inversion numerically demanding. %
\citet{MoreiraNeweySharifvaghefi24} address this difficulty by exploiting
the algebraic structure of the continuously updating GMM criterion to
compute the CU-GMM estimator.

The intuition behind the CLR test is straightforward. The AR test fixes the
null value $\beta_0$ and evaluates whether the instruments are consistent
with that value. In contrast, the CLR test compares the null to the most
favorable alternative in the entire parameter space. It asks how much larger
the rank statistic can become if we are allowed to choose the value of $%
\beta $ that best aligns the reduced form with the structural equation. If
the null is true, the rank at $\beta_0$ should already be near its maximum
and the difference in (\ref{eq:LR_statistic}) will be small. If the null is
false, there exists some $\beta$ that fits the reduced-form evidence better,
and the supremum will substantially exceed $r(\beta_0)$. By benchmarking the
null against the strongest possible alternative, the CLR typically improves
power in over-identified designs.

\paragraph{The CIL Test.}

Another robust conditional test is the CIL test proposed by %
\citet{MoreiraRidderSharifvaghefi25}. They exploit symmetries in the linear
IV model with HAC errors and construct a test based on integrating out
uncertainty via the integrated likelihood statistic
\begin{equation}
IL(\beta _{0})=\int_{-\infty }^{+\infty }\exp \left\{ \frac{1}{2}\left[
r(\beta )-r(\beta _{0})\right] \right\} |(a^{\prime }\otimes I_{k})\widehat{%
\Sigma }_{n}{}^{-1}(a\otimes I_{k})|^{-1/2}|\beta -\beta _{0}|^{k-2}d\beta ,
\label{eq:CIL}
\end{equation}%
where $a=(\beta ,1)^{\prime }$.

The intuition behind the CIL differs from that of the CLR. Instead of
comparing the null to a single best-fitting alternative, the CIL aggregates
evidence across all structural values. Each $\beta$ contributes according to
how well it aligns with the rank statistic. Alternatives that generate
larger rank statistics receive greater weight through the exponential term
in (\ref{eq:CIL}). This integration smooths local irregularities and
efficiently combines information from multiple instruments. When the model
is just-identified ($k=1$), the AR test already performs well and little is
gained from aggregation. When the model is over-identified ($k>1$), however,
the CIL can substantially outperform AR because it systematically pools
identifying information across instruments.

\paragraph{Why Inversion Becomes Difficult.}

For the CQLR test, the critical value function depends on the null only
through the one-dimensional rank statistic $r(\beta _{0})$. In contrast, the
CVFs for the CLR and CIL tests are much more complex. They may depend on the
entire vector
\begin{equation}
T(\beta _{0})=\left[ (a_{0}^{\prime }\otimes I_{k})\widehat{\Sigma }%
_{n}{}^{-1}(a_{0}\otimes I_{k})\right] ^{-1/2}(a_{0}^{\prime }\otimes I_{k})%
\widehat{\Sigma }_{n}{}^{-1}\text{vec}(R),
\end{equation}%
and directly on $\beta _{0}$ itself. Consequently, inversion no longer
reduces to the intersection of two well-behaved one-dimensional curves.
Neither the geometric threshold argument of \citet{Mikusheva10} nor the
piecewise algebraic method developed for CQLR extends directly to this
multidimensional setting.

We therefore adopt a different strategy.

\paragraph{General Approximation Framework.}

Let $\varphi(\beta_0)$ denote a general conditional test statistic and $%
c_{\alpha}(\beta_0)$ its associated critical value function. The confidence
set is defined by
\begin{equation}
\varphi(\beta_0)\leq c_{\alpha}(\beta_0).  \label{eq:CS_definition}
\end{equation}

Our strategy is to approximate both sides of (\ref{eq:CS_definition})
uniformly by functions that admit exact numerical root finding, such as
polynomials. The main obstacle is that the domain of $\beta_0$ is the entire
real line. Polynomial approximation methods provide reliable uniform
approximations only on compact sets. Direct approximation over $%
(-\infty,+\infty)$ leads to instability in the tails.

We resolve this issue by exploiting the known asymptotic behavior of the
test statistics and reparametrizing the parameter space through a smooth
bijection
\begin{equation*}
\theta _{0}=\theta (\beta _{0})\in \lbrack -1,1].
\end{equation*}%
Precise conditions on allowable compactification functions are provided in
Section \ref{section:RF model}. Because $\beta _{0}=\pm \infty $ if and only
if $\theta _{0}=\pm 1$, unbounded confidence sets can be detected by
checking whether
\begin{equation}
\overline{\varphi }(1)\leq \overline{c}_{\alpha }(1)\quad \text{or}\quad
\overline{\varphi }(-1)\leq \overline{c}_{\alpha }(-1),
\label{eq:unbounded_general}
\end{equation}%
where $\overline{\varphi }$ and $\overline{c}_{\alpha }$ denote the
reparametrized functions. In the weak-instrument environment, this feature
is essential: unbounded confidence sets occur with positive probability and
must be reported accurately.

Once mapped to the compact domain $[-1,1]$, we approximate both $\overline{%
\varphi}(\theta_0)$ and $\overline{c}_{\alpha}(\theta_0)$ using Chebyshev
polynomial interpolation. Chebyshev nodes place greater weight near the
boundaries, preventing oscillations and ensuring stable approximation even
in regions where the statistic changes rapidly. Unlike evenly spaced grids,
Chebyshev interpolation admits explicit uniform error bounds that shrink as
the approximation degree increases.

After approximating both sides of (\ref{eq:CS_definition}) by polynomials,
we solve the resulting polynomial inequality exactly to recover all boundary
points in $\theta_0$. We then map these points back to $\beta_0$ using the
inverse reparametrization $\beta_0=\theta^{-1}(\theta_0)$.

This procedure has three key advantages over grid search. First, it provides
explicit control over numerical error through the degree of approximation.
Second, it guarantees detection of unbounded confidence sets. Third, it
ensures that disconnected or narrow components are not missed due to
discretization.

Although we emphasize Chebyshev interpolation for stability and simplicity,
any uniform approximation method based on functions with exact root-finding
properties can be embedded in this framework. The central idea is to replace
arbitrary discretization with controlled uniform approximation. As the
degree of approximation increases, the coverage error induced by numerical
inversion converges to zero.

\section{Derivation of Theoretical Results}

\label{section:RF model}

In this section we formalize the statistical structure underlying the
inversion methods developed in Section \ref{section:derivation}. Our goal is
to characterize a pair of statistics that jointly govern weak-identification
robust inference under general HAC errors.

The first statistic, denoted $S(\beta _{0})$, is the component of the
continuously updating GMM objective that evaluates the structural moment
condition at the null. The second statistic, denoted $T(\beta _{0})$, is a
sufficient and complete statistic for the first-stage coefficients under the
null. Importantly, $T(\beta _{0})$ is asymptotically independent of $S(\beta
_{0})$ under standard regularity conditions. This separation allows us to
construct conditional tests whose critical value functions depend only on $%
T(\beta _{0})$.

We address two theoretical issues:

\begin{itemize}
\item The regularity conditions on the test statistic that guarantee the
parameter space can be compactified without altering the acceptance region,
thereby preventing numerical instability.

\item The way uniform approximation error propagates into coverage error,
and how this error can be made arbitrarily small by increasing the degree of
approximation.
\end{itemize}

These results provide the formal justification for the numerical procedures
introduced earlier and establish that computational tractability does not
come at the expense of statistical validity.

\subsection{Statistics and Conditional Critical Values}

To isolate the information relevant for testing the structural parameter $%
\beta $ from the nuisance parameter $\pi $ (the strength of the
instruments), it is mathematically convenient to construct a one-to-one
transformation between the reduced-form statistic $R$ and a pair of
independent random vectors. All results in this section hold exactly under
normal errors with known $\Sigma $. More generally, when $\Sigma $ is
unknown and estimated, the same arguments go through using standard
asymptotic approximations. Define
\begin{equation}
\begin{split}
S& =[(b_{0}^{\prime }\otimes I_{k})\Sigma (b_{0}\otimes
I_{k})]^{-1/2}(b_{0}^{\prime }\otimes I_{k})\mathrm{vec}(R), \\
T& =[(a_{0}^{\prime }\otimes I_{k})\Sigma ^{-1}(a_{0}\otimes
I_{k})]^{-1/2}(a_{0}^{\prime }\otimes I_{k})\Sigma {}^{-1}\mathrm{vec}(R),
\end{split}
\label{eq:S&T}
\end{equation}%
where $a_{0}=(\beta _{0},1)^{\prime }$ and $b_{0}=(1,-\beta _{0})^{\prime }$.

The statistic $S^{\prime }S$ coincides exactly with the Anderson--Rubin
statistic, while $T$ measures the strength of identification. Following %
\citet{Moreira02, Moreira09a}, the pair $(S,T)$ satisfies three key
properties:

\begin{enumerate}
\item $S$ and $T$ are statistically independent.

\item Under the null hypothesis $\beta=\beta_0$, the distribution of $S$
does not depend on any nuisance parameter.

\item $T$ is complete and sufficient for $\pi$ under the null.
\end{enumerate}

The third property is central for conditional inference. Because $T$ is
complete and sufficient for the nuisance parameter, any similar test must
also be conditionally similar given $T$. Conditioning on $T$ therefore
allows exact size control without sacrificing power. This insight underlies
the conditional approach developed by \citet{Moreira03}.

To explicitly characterize conditional rejection probabilities, we use the
inverse of the transformation in \eqref{eq:S&T}, derived by %
\citet{MoreiraMoreira19}.

\begin{lemma}
\label{lemma:inv_transf} The transformation in \eqref{eq:S&T} admits the
exact algebraic inverse
\begin{equation*}
\mathrm{vec}(R)=B(\beta _{0})S+A(\beta _{0})T,
\end{equation*}%
where
\begin{equation*}
B(\beta _{0})=\Sigma (b_{0}\otimes I_{k})[(b_{0}^{\prime }\otimes
I_{k})\Sigma (b_{0}\otimes I_{k})]^{-1/2},
\end{equation*}%
\begin{equation*}
A(\beta _{0})=(a_{0}\otimes I_{k})[(a_{0}^{\prime }\otimes I_{k})\Sigma
^{-1}(a_{0}\otimes I_{k})]^{-1/2}.
\end{equation*}
\end{lemma}

Lemma \ref{lemma:inv_transf} allows us to express any weak-IV robust test
statistic as a function of $(\beta_0,S,T)$. We therefore write $%
\varphi(\beta_0,S,T)$ interchangeably with $\varphi(\beta_0,R)$.

Under the null hypothesis and conditional on $T$, we have
\begin{equation*}
S \sim N(0,I_k).
\end{equation*}
Hence the conditional distribution of any test statistic depends only on the
known Gaussian distribution of $S$.

Let $G(y,\beta_0,T)$ denote the conditional cumulative distribution function
of $\varphi(\beta_0,S,T)$ given $T$ when $\beta=\beta_0$. The conditional
critical value function $c_\alpha(\beta_0,T)$ is defined by
\begin{equation}  \label{eq:CVF_eq}
G(c_\alpha(\beta_0,T),\beta_0,T) = 1-\alpha.
\end{equation}
This threshold leaves probability $\alpha$ in the rejection region,
conditional on the observed identification strength.

Using the Gaussian distribution of $S$, the conditional CDF can be written
explicitly as
\begin{equation*}
G(y,\beta_0,T) = \int \mathbf{1}\{\varphi(\beta_0,S,T)\le y\} \phi_k(S)\, dS,
\end{equation*}
where $\phi_k$ is the $k$-dimensional standard normal density.

For the CQLR test, the conditional distribution depends on $(\beta_0,T)$
only through the scalar statistic $T^{\prime }T$. %
\citet{AndrewsMoreiraStock07} provide an explicit representation of this
distribution, which allows computation of $c_\alpha$ by bisection between
the $(1-\alpha)$ quantiles of $\chi^2_1$ and $\chi^2_k$.

For more complex conditional tests such as CLR and CIL, analytical
evaluation of \eqref{eq:CVF_eq} is generally infeasible. In practice, we
approximate the conditional CDF by simulation: draw $S_j \sim N(0,I_k)$
independently, compute $\varphi(\beta_0,S_j,T)$ for each draw, and estimate $%
c_\alpha(\beta_0,T)$ as the empirical $(1-\alpha)$ quantile.

\subsection{Compactification and Coverage Distortions}

\label{section:general_conditional}

For general conditional tests where exact algebraic inversion is impossible,
we must uniformly approximate the inequality that defines the confidence set
and then solve the resulting polynomial inequality to obtain all distinct
interval components of the approximated set.

Traditionally, the test statistic and its CVF are treated as direct
functions of $\beta _{0}$. However, because $\beta _{0}\in (-\infty ,+\infty
)$, the parameter space is not compact. Uniform polynomial approximation
over an unbounded domain is numerically unstable and can lead to
uncontrolled oscillations in the tails. To resolve this problem permanently,
we introduce a geometric reparametrization that compactifies the parameter
space into a bounded, closed interval.

\paragraph{A Geometric Compactification.}

To motivate the compactification, consider the rank statistic $r(\beta _{0})$%
. It can be written as

\begin{equation}
\begin{aligned} r(\beta_0) &= \text{vec}(R)' \Sigma^{-1}
\left(\frac{a_0}{\|a_0\|}\otimes I_k\right) \Big[
\left(\frac{a_0'}{\|a_0\|}\otimes I_k\right) \Sigma^{-1}
\left(\frac{a_0}{\|a_0\|}\otimes I_k\right) \Big]^{-1}
\left(\frac{a_0'}{\|a_0\|}\otimes I_k\right) \Sigma^{-1} \text{vec}(R) \\ &=
\text{vec}(R)' \Sigma^{-1} (\overline a_0\otimes I_k) \Big[ (\overline
a_0'\otimes I_k) \Sigma^{-1} (\overline a_0\otimes I_k) \Big]^{-1}
(\overline a_0'\otimes I_k) \Sigma^{-1} \text{vec}(R), \end{aligned}
\end{equation}

where

\begin{equation*}
\overline a_0 := \frac{a_0}{\|a_0\|} =
\begin{bmatrix}
\beta_0/\sqrt{1+\beta_0^2} \\
1/\sqrt{1+\beta_0^2}%
\end{bmatrix}%
.
\end{equation*}

This normalization suggests a trigonometric substitution. Define

\begin{equation*}
\theta _{0}=\frac{2}{pi}\tan ^{-1}(\beta _{0})\in (-1,1),
\end{equation*}%
where $pi=3.1415926...$ This bijection maps the entire real line onto the
bounded interval $(-1,1)$. Under this transformation,

\begin{equation*}
\overline{a}_{0}=%
\begin{bmatrix}
\sin (pi.\theta _{0}/2) \\
\cos (pi.\theta _{0}/2)%
\end{bmatrix}%
.
\end{equation*}

The extreme values $\beta_0 = \pm\infty$ correspond exactly to $\theta_0 =
\pm 1$, so the compactified parameter space becomes the closed interval $%
[-1,1]$.

Let $\overline r(\theta_0)$ denote the rank statistic written as a function
of the compactified parameter.

\paragraph{Regularity of the Test Statistic.}

To ensure well-defined limits at the boundary, we impose the following mild
regularity condition. This condition is satisfied by the LR and IL test
statistics.

\begin{assumption}
\label{assump:test_stat} The test statistic $\varphi(\beta_0,R)$ is
continuous in $\beta_0$ for each fixed $R$, and admits finite limits
\begin{equation*}
\varphi_\pm(R) = \lim_{\beta_0 \to \pm\infty} \varphi(\beta_0,R).
\end{equation*}
\end{assumption}

Under Assumption \ref{assump:test_stat}, the statistic does not diverge in
the tails but instead converges to finite limits. We therefore define the
compactified test statistic

\begin{equation*}
\overline{\varphi }(\theta _{0},R)=%
\begin{cases}
\varphi (\tan (pi.\theta _{0}/2),R), & \theta _{0}\in (-1,1), \\[6pt]
\varphi _{\pm }(R), & \theta _{0}=\pm 1.%
\end{cases}%
\end{equation*}

This produces a continuous function on the compact support $[-1,1]$. The
associated CVF becomes $\overline c_\alpha(\theta_0,T)$.

\paragraph{Equivalent Characterization of the Confidence Set.}

The exact confidence set for the compactified parameter solves

\begin{equation}
\overline{\varphi}(\theta_0,R) \le \overline
c_\alpha\!\left(\theta_0,T(\theta_0,R)\right).  \label{eq:CS_general}
\end{equation}

Rather than approximating both sides of \eqref{eq:CS_general} separately, we
instead exploit the definition of the conditional CDF. Let $\overline
G(\cdot,\theta_0,T)$ denote the CDF of $\overline{\varphi}(\theta_0,R)$
conditional on $T$. By definition,

\begin{equation*}
\overline G\big(\overline c_\alpha(\theta_0,T),\theta_0,T\big) = 1-\alpha.
\end{equation*}

Therefore, Inequality \eqref{eq:CS_general} is exactly equivalent to

\begin{equation}
\overline G\!\left( \overline{\varphi}(\theta_0,R), \theta_0, T(\theta_0,R)
\right) \le 1-\alpha.  \label{eq:general_CS_CDF}
\end{equation}

Thus inversion reduces to studying the single composite function

\begin{equation*}
\theta_0 \mapsto \overline G\!\left( \overline{\varphi}(\theta_0,R),
\theta_0, T(\theta_0,R) \right).
\end{equation*}

\paragraph{Uniform Approximation and Coverage Error.}

Let $\widehat G_\varepsilon$ be a polynomial approximation satisfying

\begin{equation*}
\sup_{\theta_0 \in [-1,1]} \left| \widehat G_\varepsilon(\theta_0) -
\overline G(\theta_0) \right| \le \varepsilon.
\end{equation*}

We now quantify the coverage distortion.

\begin{proposition}
\label{prop:error_coverage_general} If $\widehat G_\varepsilon$ is a uniform
approximation with error bound $\varepsilon$, then the coverage probability
of the approximated confidence set differs from the nominal level $1-\alpha$
by at most $\varepsilon$.
\end{proposition}

\paragraph{Implications.}

Because the compactified domain $[-1,1]$ is closed and bounded, the
composite function is uniformly continuous. Standard Chebyshev approximation
theory guarantees that the uniform error $\varepsilon$ converges to zero as
the polynomial degree increases.

Therefore, the coverage distortion of the approximated confidence set
vanishes at the same rate.

This provides an explicit and transparent link between numerical
approximation error and statistical coverage, ensuring computational
reliability without sacrificing weak-identification robustness.

\section{Extensions}

\label{section:extensions}This section provides extensions from the leading
linear IV model to more general moment conditions often encountered in
applied empirical work.

\subsection{Algebraic moment conditions}

We now extend our exact algebraic approach to construct AR, LM, and CQLR
confidence sets for GMM models featuring polynomial or rational moment
conditions. These models are not theoretical curiosities. They arise in many
standard empirical settings, including dynamic panel estimators such as
Arellano-Bond moment conditions.

\paragraph{A motivating example.}

To illustrate the algebraic structure clearly, consider a simple example.
Suppose we observe $n$ iid draws $X_{i}\sim N(\mu ,1)$. Inference on $\mu $
can be based on the first two moments of the normal distribution, which
generate the simultaneous moment conditions
\begin{equation*}
\mathbb{E}[g(X_{i},\mu )]=0,\qquad g(X_{i},\mu )=%
\begin{pmatrix}
\mu -X_{i} \\
\mu ^{2}+1-X_{i}^{2}%
\end{pmatrix}%
.
\end{equation*}

The second component contains a quadratic term in $\mu$. Therefore each
component of $g(X_i,\mu)$ is a polynomial in $\mu$ of degree at most two.

The Anderson-Rubin statistic extends directly to this general GMM setting as
shown by \citet{StockWright00}. The generalized AR statistic is
\begin{equation*}
AR_n(\mu) = n\, g_n(\mu)^{\prime }W_n(\mu)^{-1} g_n(\mu),
\end{equation*}
where
\begin{equation*}
g_n(\mu) = \frac{1}{n}\sum_{i=1}^n g(X_i,\mu), \qquad W_n(\mu) = \frac{1}{n}%
\sum_{i=1}^n g(X_i,\mu) g(X_i,\mu)^{\prime }.
\end{equation*}

\paragraph{Preservation of polynomial structure.}

Because $g(X_i,\mu)$ contains terms up to degree two in $\mu$, each entry of
$W_n(\mu)$ is a polynomial of degree at most four. Indeed, each entry is a
sample average of products of two degree-two polynomials.

Writing
\begin{equation*}
W_n(\mu) =
\begin{pmatrix}
\omega_{11}(\mu) & \omega_{12}(\mu) \\
\omega_{12}(\mu) & \omega_{22}(\mu)%
\end{pmatrix}%
,
\end{equation*}
each $\omega_{ij}(\mu)$ is degree at most four. By the standard formula for
matrix inversion,
\begin{equation*}
W_n(\mu)^{-1} = \frac{1}{\omega_{11}\omega_{22}-\omega_{12}^2}
\begin{pmatrix}
\omega_{22} & -\omega_{12} \\
-\omega_{12} & \omega_{11}%
\end{pmatrix}%
.
\end{equation*}

The determinant in the denominator is a polynomial of degree at most eight.
Therefore every entry of $W_n(\mu)^{-1}$ is a rational function whose
numerator has degree at most four and whose denominator has degree at most
eight.

Since $g_n(\mu)$ is degree two in $\mu$, the quadratic form $%
g_n(\mu)^{\prime }W_n(\mu)^{-1} g_n(\mu)$ is a rational function whose
numerator and denominator are both finite-degree polynomials. In this
example, both have degree at most eight.

The crucial point is structural: quadratic forms in polynomial moments,
weighted by the inverse of their covariance matrix, preserve rationality.

\paragraph{General polynomial moment conditions.}

We now formalize this property.

Suppose the model contains $k$ moment conditions and each component of $%
g(X_i,\theta)$ is a polynomial in the scalar parameter $\theta$ of degree at
most $d$.

Then,

1. Each entry of the sample covariance matrix $W_n(\theta)$ is a polynomial
of degree at most $2d$.

2. The determinant of $W_n(\theta)$ is a polynomial of degree at most $2dk$.

3. By Cramer's rule or the Sherman-Morrison formula as demonstrated by %
\citet{MoreiraNeweySharifvaghefi24}, each entry of $W_n(\theta)^{-1}$ is a
ratio of polynomials whose numerator has degree at most $2d(k-1)$ and whose
denominator has degree at most $2dk$.

4. Since $g_n(\theta)$ is degree $d$, the AR statistic
\begin{equation*}
AR_n(\theta) = n\, g_n(\theta)^{\prime }W_n(\theta)^{-1} g_n(\theta)
\end{equation*}
is a rational function whose numerator and denominator have degree at most $%
2dk$.

Hence the AR statistic is always a rational function of finite degree.
Finite degree implies finitely many real roots of the defining polynomial
inequality, so exact confidence sets can be obtained via polynomial root
finding, exactly as in the linear IV case.

\paragraph{LM and CQLR statistics.}

The same algebraic preservation principle applies to the LM statistic.
Although its expression involves additional matrix products and inversions
(see \citet{Kleibergen05}), all building blocks are polynomial or rational
functions of $\theta$. Repeated application of the Sherman-Morrison formula
implies that the LM statistic is also a rational function of finite degree.
The maximal degree increases relative to AR but remains finite, which
guarantees exact inversion via root enumeration.

For the CQLR statistic, provided the rank statistic is itself a rational
function of $\theta$, the entire statistic becomes a composition of rational
functions\footnote{%
The expression also involves a square root, which can be eliminated by
squaring, yielding a rational representation; see Section \ref%
{section:injective_monotone_convex}}. In that case we can apply the same
partition method developed for the linear IV model: work segment by segment
where the rank statistic is monotone, identify all boundary points in the
rank domain, and map them back into $\theta$ via polynomial equations.

\paragraph{Piecewise rational moment conditions.}

Many empirical models involve piecewise polynomial or rational moment
conditions, such as models with structural breaks or threshold effects.
Within each regime the moments are polynomial or rational, so the AR and LM
statistics remain piecewise rational. We can therefore isolate each regime,
compute all boundary points within that regime using root finding, and then
assemble the global confidence set by taking unions across regimes.

\paragraph{Implication for practice.}

The central message is that polynomial GMM models inherit the same algebraic
structure as the linear IV model. Test statistics are rational functions of
finite degree. Exact confidence sets can therefore be constructed without
grid search and without numerical approximation error.

Whenever the moment conditions are polynomial or rational, confidence
regions can be computed exactly, with no risk of missing disconnected
components or truncating unbounded regions.

\subsection{Nonlinear and multivariate models}

We now consider empirical models with fully nonlinear and non-polynomial
moment conditions:
\begin{equation*}
\mathbb{E}[g(X_{i},\theta ,\gamma )]=0,
\end{equation*}%
where $\theta \in \Theta \subset \mathbb{R}^{k}$ is the structural parameter
of interest and $\gamma \in \Gamma \subset \mathbb{R}^{p}$ is a nuisance
parameter that must be profiled or partialled out.

Weak-identification robust tests extend to this setting. %
\citet{StockWright00} provide the nonlinear extension of the AR test, %
\citet{Kleibergen05} extend the LM and CQLR statistics, and %
\citet{AndrewsMikusheva16} develop the nonlinear CLR test. The inferential
logic remains unchanged: confidence sets are obtained by inverting tests
that control size under weak identification.

However, because these models lack exact polynomial structure, we cannot
rely on algebraic root-finding. Instead, we must use the uniform
approximation framework developed in Section~\ref%
{section:general_conditional}. To apply uniform approximation safely in
highly nonlinear settings, two mathematical conditions are required:
continuity and compactness.

\textbf{1. Continuity.}

First, the moment function $g(X_i,\theta,\gamma)$ must be continuous in $%
(\theta,\gamma)$. This condition is mild and satisfied in standard nonlinear
GMM applications.

Second, when the nuisance parameter $\gamma$ is profiled out by minimizing a
GMM objective function for each $\theta$, we must ensure that this
minimization step does not introduce discontinuities. This is guaranteed
under the conditions of Berge's Maximum Theorem. If the criterion function
is continuous and the nuisance parameter space $\Gamma$ is compact, then the
profiled objective remains continuous in $\theta$. This rules out sudden
jumps in the test statistic.

\textbf{2. Compact support.}

Uniform polynomial approximation requires the parameter domain to be
compact. In many applications, however, $\Theta$ and $\Gamma$ are unbounded,
for example $\mathbb{R}$ or $\mathbb{R}_+$.

This is not a fundamental obstacle. If the test statistic exhibits stable
asymptotic behavior as $(\theta,\gamma)$ diverge, the infinite domain can be
smoothly mapped onto a bounded domain without creating discontinuities. The
next proposition formalizes this idea.

\begin{enumerate}
\item
\begin{proposition}
\label{prop:nonlinear_asymptotic} Let $f:\mathbb{R}_+^{2} \to \mathbb{R}$, $%
f_{\theta}:\mathbb{R}_+ \to \mathbb{R}$, and $f_{\gamma}:\mathbb{R}_+ \to
\mathbb{R}$ be continuous functions such that

\begin{equation*}
\lim_{\theta \to \infty} f(\theta,\gamma) = f_{\theta}(\gamma) \quad \text{%
locally uniformly in } \gamma,
\end{equation*}

that is,
\begin{equation*}
\forall \varepsilon > 0,\ \forall \gamma_0,\ \exists A > 0,\ \exists \delta
> 0 \text{ such that } \theta > A,\ \gamma \in (\gamma_0 - \delta,\gamma_0 +
\delta) \implies |f(\theta,\gamma) - f_{\theta}(\gamma)| < \varepsilon,
\end{equation*}

and

\begin{equation*}
\lim_{\gamma \to \infty} f(\theta,\gamma) = f_{\gamma}(\theta) \quad \text{%
locally uniformly in } \theta,
\end{equation*}

with the compatibility condition

\begin{equation*}
\lim_{\theta \to \infty} f_{\gamma}(\theta) = \lim_{\gamma \to \infty}
f_{\theta}(\gamma) = L.
\end{equation*}

Define the compactified function $g:[0,1]^2 \to \mathbb{R}$ by

\begin{equation*}
g(x,y)=
\begin{cases}
f(\tan (pi \cdot x/2),\tan (pi \cdot y/2)) & \text{if } x,y \neq 1, \\
f_{\gamma}(\tan (pi \cdot x/2)) & \text{if } x \neq 1,\ y=1, \\
f_{\theta}(\tan (pi \cdot y/2)) & \text{if } x=1,\ y \neq 1, \\
L & \text{if } x=y=1.%
\end{cases}%
\end{equation*}

Then $g$ is continuous on $[0,1]^2$.
\end{proposition}
\end{enumerate}

This proposition formalizes a practical principle. If the test statistic
flattens out and converges smoothly to well-defined limits when parameters
diverge, then a trigonometric transformation maps the infinite parameter
space onto a bounded domain while preserving continuity.

Once continuity on a compact domain is established, we approximate the
nonlinear test inequality uniformly by a polynomial inequality. By
Proposition~\ref{prop:error_coverage_general}, a uniform bound on
approximation error implies a strict bound on coverage distortion. The
resulting confidence set is semi-algebraic, meaning it is defined by
finitely many polynomial equalities and inequalities.

In the one-dimensional case, all boundary points are roots of a single
polynomial, which can be computed rapidly using standard numerical
eigenvalue routines such as the QR algorithm.

When $\theta$ is multidimensional, the confidence region may have a complex
geometry. Nevertheless, it remains semi-algebraic. Cylindrical Algebraic
Decomposition (CAD) decomposes such sets into finitely many simple cells. By
the Tarski-Seidenberg theorem, projections of semi-algebraic sets remain
semi-algebraic. Hence even after projecting onto a lower-dimensional
parameter of interest, the bounds of the confidence set can still be
computed using polynomial root-finding.

The main implication for empirical practice is that the uniform
approximation framework developed for the linear IV model extends naturally
to nonlinear GMM models. Under mild continuity and asymptotic regularity
conditions, researchers can construct confidence sets with arbitrarily small
coverage distortion, avoiding the numerical failures and hidden omissions
that arise from standard grid search methods.

\section{Conclusion}

\label{section:conclusion}

This paper develops new methods for constructing confidence sets for
structural parameters in linear IV models when instruments may be weak and
errors may be heteroskedastic, autocorrelated, or clustered. While the
literature has established tests that remain valid under weak
identification, empirical practice typically relies on grid search to invert
those tests. We show that this numerical step is not innocuous. Grid
procedures often miss disconnected components, truncate unbounded regions,
and generate confidence sets that are wider or qualitatively different from
the true acceptance region. These distortions arise from arbitrary
discretization choices rather than from the underlying statistical theory.

Our approach replaces grid inversion with exact and approximation-based
methods that respect the algebraic structure of the test statistics. For the
AR and LM tests, we exploit their rational form to characterize the
confidence set as the solution to a polynomial inequality and recover all
boundary points via polynomial root finding. The same logic extends to
models with polynomial or rational moment conditions, since the algebraic
structure of the statistics is preserved.

For the CQLR test, we use the geometry of the statistic and the monotonicity
and convexity properties of the critical value function to derive an exact
inversion algorithm. For more general conditional tests, including CLR and
CIL, we construct uniform polynomial approximations to the inequality that
defines the confidence set. The approximation error for the test statistic
translates directly into coverage error, and both can be made arbitrarily
small by increasing the degree of approximation.

Across a wide range of empirical specifications drawn from the literature,
our exact and approximation-based methods reliably recover the true
confidence sets, while grid search frequently fails. The discrepancies are
particularly pronounced in designs where the confidence region contains
narrow components, exhibits sharp curvature, or extends far into the tails.
In these cases, coarse grids can materially distort inference.

The methods developed here are straightforward to implement and apply to a
broad class of models. They provide researchers with practical tools for
reporting confidence sets that remain valid under weak identification and
complex error structures, without relying on arbitrary numerical choices.
More broadly, the results highlight that numerical inversion is not a minor
computational detail but a central component of valid weak-identification
robust inference. By replacing grid search with algebraically grounded
procedures, we offer a transparent and theoretically disciplined framework
for empirical practice.

\bibliographystyle{plainnat}
\bibliography{References_fixed}

\newpage

\section{CQLR Confidence Region and Exact Inversion Algorithm}

\label{appendix:cqlr_exact_inversion}

To implement exact inversion under HAC errors, we must address two primary
difficulties: (i) the mapping $r(\beta_0)$ is only piecewise invertible in $%
\beta_0$, and (ii) the QLR statistic may change monotonicity or curvature
within each piece. Exact inversion therefore proceeds by first decomposing
the parameter space into regions where the geometry is well behaved, and
then locating all intersections in the rank domain before mapping back to $%
\beta_0$.

Formally, this chapter executes the following sequence:

\begin{enumerate}
\item \textbf{Partitioning:} We partition the parameter space into maximal
intervals on which $r(\beta)$ is injective, allowing us to define the test
statistic implicitly as a function of $r$.

\item \textbf{Mapping the Geometry:} Within each interval, we determine all
points where the QLR statistic changes monotonicity or curvature as a
function of the rank statistic.

\item \textbf{Root Finding:} We apply a taylored root-finding procedure to
compute all solutions to $QLR(r) \leq \kappa_\alpha(r)$ inside each valid
interval.

\item \textbf{Reconstruction:} After collecting the valid intervals for the
rank statistic, we map them back to the structural parameter $\beta_0$ using
the algebraic structure of the rank statistic.
\end{enumerate}

\subsection{Finding All Injective, Monotonicity, and Convexity Intervals}

\label{section:injective_monotone_convex}

Our first job is to write the QLR statistic as a function of $r$. Because we
can only do this over intervals where the function $\beta\mapsto r(\beta)$
is injective, we must locate these intervals and subsequently determine the
shape of the test statistic within them.

\subsubsection*{Injective Intervals}

Since the rank statistic is rational, its derivative is also rational, so we
are able to numerically find all the real roots $\beta_1<\cdots<\beta_{m-1}$
of $r^{\prime}(\beta)$. Defining $\beta_0=-\infty$, $\beta_m=+\infty$, the
function $r(\beta)$ is injective on $[\beta_{i-1},\beta_{i}]$, for $%
i=1,\cdots,m$.

For each of these intervals, we can define the inverse function $\beta^i(r)$
and write the test statistic as an implicit function of $r$:
\begin{equation*}
g_i:r\in[r^i_0,r^i_1]\mapsto QLR(\beta(r))
\end{equation*}
where the boundaries in the rank domain are:
\begin{equation*}
r^i_0=\inf_{\beta\in[\beta_{i-1},\beta_i]}r(\beta)=\min\{r(\beta_{i-1}),r(%
\beta_i)\}
\end{equation*}
\begin{equation*}
r^i_1=\sup_{\beta\in[\beta_{i-1},\beta_i]}r(\beta)=\max\{r(\beta_{i-1}),r(%
\beta_i)\}
\end{equation*}

\subsubsection*{Monotonicity and Convexity Intervals}

We now provide an algorithm to find increasing/decreasing and convex/concave
intervals for the function $g_i$ on $[r^i_{0},r^i_{1}]$, by finding all the
roots of $g_i^{\prime }$ and $g_i^{\prime \prime }$. To simplify notation,
define:
\begin{equation*}
\Delta(\beta):=(AR(\beta)-r(\beta))^2+4LM(\beta)\cdot r(\beta)
\end{equation*}
By the Chain Rule, the first and second derivatives of $g_i$ are given by:
\begin{equation*}
g_i^{\prime }(r)=QLR^{\prime }(\beta^i(r))\cdot (\beta^i)^{\prime }(r)
\end{equation*}
\begin{equation*}
g_i^{\prime \prime }(r)=Q^{\prime \prime }(\beta^i(r))\cdot
((\beta^i)^{\prime}(r))^2+Q^{\prime }(\beta^i(r))\cdot (\beta^i)^{\prime
\prime }(r)
\end{equation*}
Using the Inverse Mapping Theorem to find $(\beta^i)^{\prime }(\cdot)$ and $%
(\beta^i)^{\prime \prime }(\cdot)$, and setting the derivatives of $g_i$ to
zero, we get the following equations in terms of the structural parameter $%
\beta$
\begin{equation}  \label{eq:g_der}
\Delta(\beta)^{\prime}+2\Delta(\beta)^{1/2}(AR(\beta)-r(\beta))^{\prime}=0
\end{equation}

\begin{equation}  \label{eq:g_der_2}
\begin{aligned} 4\Delta(\beta)^{3/2}\left[(AR(\beta)-r(\beta))^{\prime
\prime }r(\beta)^{\prime }-(AR(\beta)-r(\beta))^{\prime }r(\beta)^{\prime
\prime }\right]+\\ +\left[2\Delta(\beta)\Delta(\beta)^{\prime \prime
}-(\Delta(\beta)^{\prime })^2\right]r(\beta)^{\prime
}-2\Delta(\beta)\Delta(\beta)^{\prime }r(\beta)^{\prime \prime }=0
\end{aligned}
\end{equation}

Since $AR(\cdot)$, $r(\cdot)$ and $\Delta(\cdot)$ are rational functions, we
can isolate the term $\Delta(\beta)^{1/2}$ in equations \ref{eq:g_der} and %
\ref{eq:g_der_2} and take the square to get rational equations in $\beta$,
that we can solve numerically. Although taking the square of the equations
may introduce new roots, we can rule them out by checking if indeed
derivatives change sign at these points. This step also rules out points of
inflection.

\subsection{Root Finding Algorithm and Mapping Back to the Structural
Parameter}

\label{section:root_finding_mapping}

Let $\{\beta_j\}_{j\in J}$ be the union of all $\beta$'s found with the
procedures in Section \ref{section:injective_monotone_convex}. The intervals
generated by these points are such that the test statistic can be implicitly
defined as a functions of $r$, and these functions change neither their
monotonicity nor their convexity on these intervals.

For each interval $[\beta_0,\beta_1]$ generated by these points, we view the
test statistic $g$ as a function of $r\in[r_0,r_1]$. We already know $%
\kappa(r)$ is strictly decreasing and convex. We evaluate the roots of $g(r)
= \kappa(r)$ based on the geometric behavior of $g$ on $[r_0,r_1]$.

\subsubsection*{Case A: $g$ is Increasing}

In this case, there is at most one $\overline{r }\in [r_{0},r_{1}]$ such
that $g(\overline{r })=\kappa (\overline{r })$.

\begin{itemize}
\item If $g(r_{0})>\kappa (r_{0})$ or $g(r_{1})<\kappa (r_{1})$, there is no
such $\overline{r}$.

\item If $g(r_{0})=\kappa (r_{0})$ (or $g(r_{1})=\kappa (r_{1})$), then $%
\overline{r }=r_{0}$ (or $\overline{r }=r_{1}$).

\item If $g(r_{0})<\kappa (r_{0})$ and $g(r_{1})>\kappa (r_{1})$, then $%
\overline{r }\in (r_0,r_1)$ is the only root of $\kappa -g$, and it can be
found using the Standard Bisection Method.
\end{itemize}

\subsubsection*{Case B: $g$ is Decreasing and Concave}

\begin{itemize}
\item If $g(r_{0})\geq \kappa (r_{0})$ and $g(r_{1})\geq \kappa (r_{1})$,
then there is no intersection point on $(r_{0},r_{1})$.

\item If $g(r_{0})<\kappa (r_{0})$ and $g(r_{1})\geq \kappa(r_{1})$ \textbf{%
OR} if $g(r_{0})\geq \kappa (r_{0})$ and $g(r_{1})<\kappa(r_{1})$, there is
exactly one intersection point $\overline{r }$ on $(r_{0},r_{1})$, which can
be found via the \textbf{Generalized Bisection Method} for $\kappa -g$.

\item If $g(r_{0})<\kappa (r_{0})$ and $g(r_{1})<\kappa (r_{1})$, we can
have zero, one, or two intersection points. The function $h:=\kappa -g$ is
strictly convex, so we can numerically find its minimum $M$ and unique
minimizer $r^{\ast }\in [r_{0},r_{1}]$\footnote{%
Here we can use standard global algorithms for convex optimization}.

\begin{itemize}
\item If $M>0$, there is no intersection point on $[r_{0},r_{1}]$.

\item If $M=0$, the only intersection point is $r^{\ast }$.

\item If $M<0$, there are two intersection points: one in $(r_{0},r^{\ast })$
and another in $(r^{\ast },r_{1})$. Since $h(r_{0}),h(r_{1})>0$ and $%
h(r^{\ast })<0$, we find both via the Standard Bisection Method.
\end{itemize}
\end{itemize}

\subsubsection*{Case C: $g$ is Decreasing and Convex}

This is the most difficult case to handle. Finding intersection points
requires \textbf{Algorithm 1}, introduced below, which systematically
recovers all solutions. In order to intuitively understand the mechanism
behind this algorithm, let us consider a particular example where $g$ is
linear, which is the case under homoskedasticity (see Figure \ref%
{fig:algorithm1}). You can see the graphs of $g$ and $\kappa $ intercept
each other at two points.

In order to find the first intersection point, we draw a red line starting
from $(r _{0},g(r _{0}))$ with the largest slope in such a way the whole
line is bellow the graph of $\kappa $, and we determine $r ^{\ast }$, the
point at which the red line and the graph of $\kappa $ are tangent. Because
of the convexity and monotonicity of both $\kappa$ and $g$, we ensure there
is at most one intersection point between $r_0$ and the tangency point $%
r^{\ast}$. Once found, we restart the algorithm for the remainder of the
interval with $r^*$ as the updated value for the lower bound $r_0$.

\begin{figure}[h]
\caption{Illustration of Algorithm 1}
\label{fig:algorithm1}\centering
{\normalsize \centering
\includegraphics[width=0.8\linewidth]{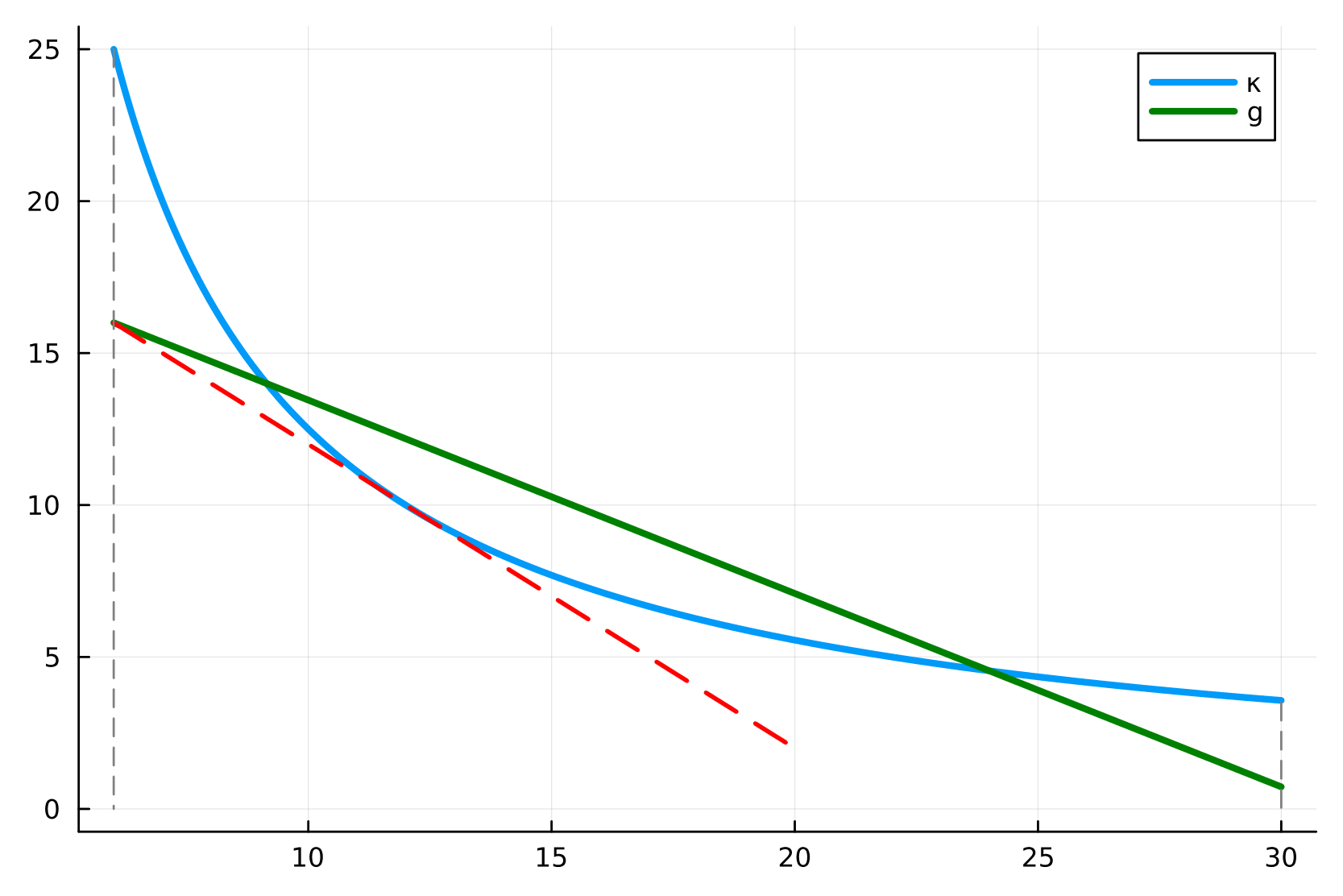} }
\end{figure}

Our algorithm uses the following generalization of the bisection method for
monotone, but not necessarily strictly monotone, functions.

\vspace{0.5cm} \noindent\textbf{Generalized Bisection Method:}\newline
Suppose we know the function $h(x)$ is strictly positive for $x<x^*$ and
non-positive for $x\geq x^*$, then we can use the following algorithm to
find $x^*\in(a,b)$ with a tolerance $\varepsilon>0$:

\begin{enumerate}
\item Define $x_0=a$, $x_1=b$.

\item Take the midpoint $\overline{x}=(x_1+x_0)/2$.

\item If $x_1-x_0\leq \varepsilon$, return $x^*\approx \overline{x}$.

\item Otherwise, if $h(\overline{x})>0$ ($h(\overline{x})\leq0$), redefine $%
x_0=\overline{x}$ ($x_1=\overline{x}$) and go back to step 2.
\end{enumerate}

A symmetric algorithm applies if $h$ is strictly positive after $x^*$ and
non-positive before $x^*$.

\vspace{0.5cm} \noindent\textbf{Algorithm 1:} Here we provide a step-by-step
procedure to find all the roots of the equation $\kappa-g=0$ over $[r_0,r_1]$%
, when $g$ in decreasing and convex. Proposition \ref{prop:algorithm1_step1}
in Appendix \ref{section:proofs} formalizes the claims presented in the
description of this algorithm. We have five possibilities of ordering for
the values of the functions at $r_0$ and $r_1$:

\begin{enumerate}
\item \textbf{If $g(r_{0})<\kappa (r_{0})$}: Define the nonincreasing
function
\begin{equation*}
H(r )=\kappa (r )-g(r_{0})-\kappa ^{\prime }(r )(r -r_{0})
\end{equation*}
and
\begin{equation*}
r^{\ast }=\sup \{r \in [r_{0},r_{1}]:H(r )>0\}
\end{equation*}
Here $r^*$ is exactly the point where the red line is tangent to the graph
of $\kappa$ in Figure \ref{fig:algorithm1}. Since $H(r_{0})>0$ and $%
H^{\prime }(r )\leq 0$, if $H(r_{1})>0$ then $r^{\ast }=r_{1}$. Otherwise,
find $r^{\ast }$ via the Generalized Bisection Method.

\begin{enumerate}
\item If $g(r^{\ast })<\kappa (r^{\ast })$, there is no intersection point
on $[r_{0},r^{\ast }]$. Set $r_{0}=r^{\ast }$ and repeat Step 1.

\item If $g(r^{\ast })>\kappa (r^{\ast })$, exactly one intersection exists
on $(r_{0},r^{\ast })$, and it can be found using the Standard Bisection
Method. Set $r_{0}=r^{\ast }$ and continue the algorithm with \textbf{Case 2}%
.

\item If $g(r^{\ast })=\kappa (r^{\ast })$, the only intersection on $%
[r_{0},r^{\ast }]$ is $r^{\ast }$. Set $r_{0}=r^{\ast }$ and continue the
algorithm with: \textbf{Case 3} if $g(r_1)<\kappa(r_1)$; \textbf{Case 4} if $%
g(r_1)>\kappa(r_1)$; \textbf{Case 5} if $g(r_1)=\kappa(r_1)$.
\end{enumerate}

\item \textbf{If $g(r_{0})>\kappa (r_{0})$}: Follow \textbf{Case 1}, but
interchange the roles of $\kappa $ and $g$.

\item \textbf{If $g(r_{0})=\kappa (r_{0})$ and $g(r_{1})<\kappa(r_{1})$}:
Apply a backward version of \textbf{Case 1}. Define the nondecreasing
function
\begin{equation*}
H(r )=\kappa (r )-g(r_{1})-\kappa ^{\prime }(r )(r -r_{1})
\end{equation*}
and
\begin{equation*}
r^{\ast }=\inf \{r \in [r_{0},r_{1}]:H(r )>0\}
\end{equation*}
If $H(r_{0})>0$, then $r^{\ast }=r_{0}$. Otherwise, find $r^{\ast }$ via
Generalized Bisection.

\begin{enumerate}
\item If $g(r^{\ast })<\kappa (r^{\ast })$, there are no intersection points
on $[r^{\ast }, r_{1}]$. Set $r_{1}=r^{\ast }$ and continue with \textbf{%
Case 3}.

\item If $g(r^{\ast })>\kappa (r^{\ast })$, there is exactly one
intersection on $(r^{\ast },r_{1})$, that can be found using Standard
Bisection. Set $r_{1}=r^{\ast }$ and continue with \textbf{Case 3}, now
interchanging the roles of $\kappa $ and $g$.

\item If $g(r^{\ast })=\kappa (r^{\ast })$, the only intersection on $%
[r^{\ast },r_{1}]$ is $r^{\ast }$. Set $r_{1}=r^{\ast }$ and continue with
\textbf{Case 5}.
\end{enumerate}

\item \textbf{If $g(r_{0})=\kappa (r_{0})$ and $g(r_{1})>\kappa (r_{1})$}:
Follow \textbf{Case 3}, interchanging the roles of $\kappa $ and $g$.

\item \textbf{If $g(r_{0})=\kappa (r_{0})$ and $g(r_{1})=\kappa (r_{1})$}:
Take $\overline{r}$ as the midpoint of the interval $[r_{0},r_1]$, and
continue the algorithm for each sub-interval $[r_{0},\overline{r }]$ and $[%
\overline{r },r_{1}]$. The case we follow for each sub-interval depends on
the ordering between $g$ and $\kappa$ at the endpoints.
\end{enumerate}

The algorithm stops if $r_{1}-r_{0} \leq \varepsilon$.

\subsubsection*{Mapping Back to the Structural Parameter}

For each solution $r^*$ of equation $g(r)=\kappa (r)$ on the interval $%
[r_{0},r_{1}]$, we map them back to the structural parameter $\beta$ using
the rational representation of the rank statistic. We solve the equation
\begin{equation*}
r(\beta)=r^*
\end{equation*}
for $\beta\in[\beta_0,\beta_1]$ numerically using polynomial root finding
algorithms.

\newpage

\section{Additional Results}

{\normalsize \label{section:proofs} }

\begin{proposition}
\label{prop:algorithm1_step1} Let $\kappa$ be the CVF of the CQLR test, $g$
be a convex and decreasing function over $[r_0,r_1]$, such that $%
g(r_0)<\kappa(r_0)$, and define
\begin{equation*}
H(r)=\kappa(r)-g(r_0)-\kappa^{\prime }(r)(r-r_0)
\end{equation*}
\begin{equation*}
r^*=\sup\{r\in[r_0,r_1]:H(r)>0\}
\end{equation*}
Then we have the following:

\begin{enumerate}
\item If $g(r^{\ast })<\kappa (r^{\ast })$, there is no intersection point
on the interval $[r_{0},r^{\ast }]$.

\item If $g(r^{\ast })>\kappa (r^{\ast })$, then there exists exactly one
intersection point on the interval $(r_{0},r^{\ast })$.

\item If $g(r^{\ast })=\kappa (r^{\ast })$, then the only intersection point
on the interval $[r_{0},r^{\ast }]$ is $r^{\ast }$.
\end{enumerate}
\end{proposition}

\newpage

\section{Additional figures and tables}

{\normalsize \label{appendix:figs_tables} }

\subsection{Appendix for Section \protect\ref{section:emp_relevance}}

\label{supp:emp_relevance}

For completeness, we revisit the empirical application from Section \ref%
{section:emp_relevance}, using stock returns rather than interest rates as
the endogenous variable. Because stock returns are less predictable, weak
identification is more severe. Consequently, Table \ref{tab:EIS_stock} shows
that weak-IV robust tests usually yield unbounded sets, unlike the
mechanically bounded t-ratio. Furthermore, Table \ref{tab:compare_stock}
confirms our earlier results: grid confidence sets are wider and can miss
interval components, as demonstrated by the LM test for Canada.

{\normalsize
\begin{table}[!h]
{\normalsize
\scalebox{0.92}{
\centering
\begin{threeparttable}
\caption{Confidence Intervals for the EIS using Stock Returns}
\label{tab:EIS_stock}
\begin{tabular}{cccccccc}
\toprule
 & Robust & t-ratio & AR & LM & CQLR & CLR & CIL \\ \midrule
\multirow{2}{*}{AUL} & Yes & {[}-0.02, 0.12{]} & {[}$-\infty, +\infty${]} & {[}$-\infty, +\infty${]} & {[}$-\infty, +\infty${]} & {[}$-\infty, +\infty${]} & {[}$-\infty, +\infty${]} \\
 & No & {[}-0.02, 0.12{]} & {[}$-\infty, +\infty${]} & {[}$-\infty, +\infty${]} & {[}$-\infty, +\infty${]} & {[}$-\infty, +\infty${]} & {[}$-\infty, +\infty${]} \\ \midrule
\multirow{2}{*}{CAN} & Yes & {[}0.00, 0.24{]} & {[}$-\infty, +\infty${]} & {[}-0.10, 0.49{]} & {[}0.04, 0.63{]} & {[}0.04, 0.67{]} & {[}0.05, 0.71{]} \\
 & No & {[}0.03, 0.22{]} & {[}0.02, 2.28{]} & {[}-0.11, 0.33{]} & {[}0.05, 0.39{]} & {[}0.05, 0.38{]} & {[}0.05, 0.41{]} \\ \midrule
\multirow{2}{*}{FRA} & Yes & {[}-0.08, 0.04{]} & {[}-0.27, 0.06{]} & {[}-0.11, 0.31{]} & {[}-0.13, 0.04{]} & {[}-0.14, 0.03{]} & {[}-0.16, 0.05{]} \\
 & No & {[}-0.09, 0.05{]} & {[}-0.25, 0.18{]} & {[}$-\infty, +\infty${]} & {[}-0.15, 0.10{]} & {[}-0.15, 0.10{]} & {[}$-\infty, +\infty${]} \\ \midrule
\multirow{2}{*}{GER} & Yes & {[}-0.18, 0.13{]} & {[}$-\infty, +\infty${]} & {[}$-\infty, +\infty${]} & {[}$-\infty, +\infty${]} & {[}$-\infty, +\infty${]} & {[}$-\infty, +\infty${]} \\
 & No & {[}-0.16, 0.11{]} & {[}$-\infty, +\infty${]} & {[}$-\infty, +\infty${]} & {[}$-\infty, +\infty${]} & {[}$-\infty, +\infty${]} & {[}$-\infty, +\infty${]} \\ \midrule
\multirow{2}{*}{ITA} & Yes & {[}-0.05, 0.07{]} & {[}$-\infty, +\infty${]} & {[}$-\infty, +\infty${]} & {[}$-\infty, +\infty${]} & {[}$-\infty, +\infty${]} & {[}$-\infty, +\infty${]} \\
 & No & {[}-0.05, 0.06{]} & {[}$-\infty, +\infty${]} & {[}$-\infty, +\infty${]} & {[}$-\infty, +\infty${]} & {[}$-\infty, +\infty${]} & {[}$-\infty, +\infty${]} \\ \midrule
\multirow{2}{*}{JAP} & Yes & {[}-0.02, 0.12{]} & {[}-0.04, 0.21{]} & {[}$-\infty, +\infty${]} & {[}-0.02, 0.17{]} & {[}-0.02, 0.16{]} & {[}-0.02, 0.16{]} \\
 & No & {[}-0.01, 0.12{]} & {[}-0.04, 0.30{]} & {[}-0.94, 0.19{]} & {[}-0.02, 0.20{]} & {[}-0.02, 0.20{]} & {[}-0.01, 0.19{]} \\ \midrule
\multirow{2}{*}{NTH} & Yes & {[}-0.14, 0.20{]} & {[}$-\infty, +\infty${]} & {[}$-\infty, +\infty${]} & {[}$-\infty, +\infty${]} & {[}$-\infty, +\infty${]} & {[}$-\infty, +\infty${]} \\
 & No & {[}-0.13, 0.19{]} & {[}$-\infty, +\infty${]} & {[}$-\infty, +\infty${]} & {[}$-\infty, +\infty${]} & {[}$-\infty, +\infty${]} & {[}$-\infty, +\infty${]} \\ \midrule
\multirow{2}{*}{SWE} & Yes & {[}-0.06, 0.03{]} & {[}$-\infty, +\infty${]} & {[}$-\infty, +\infty${]} & {[}$-\infty, +\infty${]} & {[}$-\infty, +\infty${]} & {[}$-\infty, +\infty${]} \\
 & No & {[}-0.06, 0.04{]} & {[}$-\infty, +\infty${]} & {[}$-\infty, +\infty${]} & {[}$-\infty, +\infty${]} & {[}$-\infty, +\infty${]} & {[}$-\infty, +\infty${]} \\ \midrule
\multirow{2}{*}{SWI} & Yes & {[}-0.35, 0.25{]} & {[}$-\infty, +\infty${]} & {[}$-\infty, +\infty${]} & {[}$-\infty, +\infty${]} & {[}$-\infty, +\infty${]} & {[}$-\infty, +\infty${]} \\
 & No & {[}-0.41, 0.32{]} & {[}$-\infty, +\infty${]} & {[}$-\infty, +\infty${]} & {[}$-\infty, +\infty${]} & {[}$-\infty, +\infty${]} & {[}$-\infty, +\infty${]} \\ \midrule
\multirow{2}{*}{UK} & Yes & {[}-0.09, 0.07{]} & {[}$-\infty, +\infty${]} & {[}$-\infty, +\infty${]} & {[}$-\infty, +\infty${]} & {[}$-\infty, +\infty${]} & {[}$-\infty, +\infty${]} \\
 & No & {[}-0.08, 0.06{]} & {[}-0.33, -0.03{]} & {[}$-\infty, +\infty${]} & {[}$-\infty, +\infty${]} & {[}$-\infty, +\infty${]} & {[}$-\infty, +\infty${]} \\ \midrule
\multirow{2}{*}{USA} & Yes & {[}-0.00, 0.06{]} & {[}$-\infty, +\infty${]} & {[}$-\infty, +\infty${]} & {[}$-\infty, +\infty${]} & {[}$-\infty, +\infty${]} & {[}$-\infty, +\infty${]} \\
 & No & {[}-0.00, 0.07{]} & {[}$-\infty, +\infty${]} & {[}$-\infty, +\infty${]} & {[}$-\infty, +\infty${]} & {[}$-\infty, +\infty${]} & {[}$-\infty, +\infty${]} \\ \bottomrule
\end{tabular}
\begin{tablenotes}\footnotesize
\item Column "Robust" indicate whether the CS is robust to heteroskedasticity or not. $\emptyset$ indicates an empty CS.
\item CIs are calculated using our methodology
\end{tablenotes}
\end{threeparttable}
}  }
\end{table}
}

{\normalsize
\begin{table}[h]
{\normalsize
\scalebox{0.85}{
\centering
\begin{threeparttable}
\caption{Comparison of Yogo's Reports and Our Method (Stock Returns)}
\label{tab:compare_stock}
\begin{tabular}{cccccccc}
\toprule
 & Robust & Yogo - AR & AR & Yogo - LM & LM & Yogo - CLR & CLR \\ \midrule
\multirow{2}{*}{AUL} & Yes &  & {[}$-\infty, +\infty${]} &  & {[}$-\infty, +\infty${]} &  & {[}$-\infty, +\infty${]} \\
 & No & {[}$-\infty, +\infty${]} & {[}$-\infty, +\infty${]} & {[}$-\infty, +\infty${]} & {[}$-\infty, +\infty${]} & {[}$-\infty, +\infty${]} & {[}$-\infty, +\infty${]} \\ \midrule
\multirow{2}{*}{CAN} & Yes &  & {[}$-\infty, +\infty${]} &  & {[}-0.10, 0.49{]} &  & {[}0.04, 0.67{]} \\
 & No & {[}0.02, 4.03{]} & {[}0.02, 2.28{]} & {[}0.05, 0.35{]} & {[}-0.11, 0.33{]} & {[}0.04, 0.41{]} & {[}0.05, 0.38{]} \\ \midrule
\multirow{2}{*}{FRA} & Yes &  & {[}-0.27, 0.06{]} &  & {[}-0.11, 0.31{]} &  & {[}-0.14, 0.03{]} \\
 & No & {[}-0.28, 0.20{]} & {[}-0.25, 0.18{]} & {[}$-\infty, +\infty${]} & {[}$-\infty, +\infty${]} & {[}-0.16, 0.11{]} & {[}-0.15, 0.10{]} \\ \midrule
\multirow{2}{*}{GER} & Yes &  & {[}$-\infty, +\infty${]} &  & {[}$-\infty, +\infty${]} &  & {[}$-\infty, +\infty${]} \\
 & No & {[}$-\infty, +\infty${]} & {[}$-\infty, +\infty${]} & {[}$-\infty, +\infty${]} & {[}$-\infty, +\infty${]} & {[}$-\infty, +\infty${]} & {[}$-\infty, +\infty${]} \\ \midrule
\multirow{2}{*}{ITA} & Yes &  & {[}$-\infty, +\infty${]} &  & {[}$-\infty, +\infty${]} &  & {[}$-\infty, +\infty${]} \\
 & No & {[}$-\infty, +\infty${]} & {[}$-\infty, +\infty${]} & {[}$-\infty, +\infty${]} & {[}$-\infty, +\infty${]} & {[}$-\infty, +\infty${]} & {[}$-\infty, +\infty${]} \\ \midrule
\multirow{2}{*}{JAP} & Yes &  & {[}-0.04, 0.21{]} &  & {[}$-\infty, +\infty${]} &  & {[}-0.02, 0.16{]} \\
 & No & {[}-0.05, 0.32{]} & {[}-0.04, 0.30{]} & {[}-1.01, 0.20{]} & {[}-0.94, 0.19{]} & {[}-0.02, 0.21{]} & {[}-0.02, 0.20{]} \\ \midrule
\multirow{2}{*}{NTH} & Yes &  & {[}$-\infty, +\infty${]} &  & {[}$-\infty, +\infty${]} &  & {[}$-\infty, +\infty${]} \\
 & No & {[}$-\infty, +\infty${]} & {[}$-\infty, +\infty${]} & {[}$-\infty, +\infty${]} & {[}$-\infty, +\infty${]} & {[}$-\infty, +\infty${]} & {[}$-\infty, +\infty${]} \\ \midrule
\multirow{2}{*}{SWE} & Yes &  & {[}$-\infty, +\infty${]} &  & {[}$-\infty, +\infty${]} &  & {[}$-\infty, +\infty${]} \\
 & No & {[}$-\infty, +\infty${]} & {[}$-\infty, +\infty${]} & {[}$-\infty, +\infty${]} & {[}$-\infty, +\infty${]} & {[}$-\infty, +\infty${]} & {[}$-\infty, +\infty${]} \\ \midrule
\multirow{2}{*}{SWI} & Yes &  & {[}$-\infty, +\infty${]} &  & {[}$-\infty, +\infty${]} &  & {[}$-\infty, +\infty${]} \\
 & No & {[}$-\infty, +\infty${]} & {[}$-\infty, +\infty${]} & {[}$-\infty, +\infty${]} & {[}$-\infty, +\infty${]} & {[}$-\infty, +\infty${]} & {[}$-\infty, +\infty${]} \\ \midrule
\multirow{2}{*}{UK} & Yes &  & {[}$-\infty, +\infty${]} &  & {[}$-\infty, +\infty${]} &  & {[}$-\infty, +\infty${]} \\
 & No & {[}-0.51, -0.02{]} & {[}-0.33, -0.03{]} & {[}$-\infty, +\infty${]} & {[}$-\infty, +\infty${]} & {[}$-\infty, +\infty${]} & {[}$-\infty, +\infty${]} \\ \midrule
\multirow{2}{*}{USA} & Yes &  & {[}$-\infty, +\infty${]} &  & {[}$-\infty, +\infty${]} &  & {[}$-\infty, +\infty${]} \\
 & No & {[}$-\infty, +\infty${]} & {[}$-\infty, +\infty${]} & {[}$-\infty, +\infty${]} & {[}$-\infty, +\infty${]} & {[}$-\infty, +\infty${]} & {[}$-\infty, +\infty${]} \\ \bottomrule
\end{tabular}
\begin{tablenotes}\footnotesize
\item Column "Robust" indicate whether the CS is robust to heteroskedasticity or not. $\emptyset$ indicates an empty CS.
\end{tablenotes}
\end{threeparttable}
}  }
\end{table}
}

\newpage

\subsection{Appendix for Section \protect\ref{section:CQLR}}

\label{supp:CQLR}

\citet{AndrewsMoreiraStock07} provide an explicit formula for the
conditional CDF of the QLR statistic:
\begin{equation*}
G(x,r):=\mathbb{P}(QLR(r)\leq x|r)=\int_0^1F_k\left(\dfrac{x(x+r)}{x+ru^2}%
\right)(1-u^2)^{(k-3)/2}du
\end{equation*}
The CVF, $\kappa_\alpha(r)$, is implicitly defined by the integral equation:
\begin{equation*}
G(\kappa_\alpha(r),r)=1-\alpha.
\end{equation*}

We obtain formulas for its derivatives via implicit differentiation. Because
the conditional CDF does not admit a closed-form expression, the integrals
involved in $G(x,r)$ and its derivatives are evaluated numerically.

We verify the shape of the CVF numerically. In particular, for $%
k=1,\ldots,150$, we find that the first derivative of $\kappa_\alpha(r)$ is
strictly negative and the second derivative is strictly positive over the
relevant domain. These properties confirm that the CVF is strictly
decreasing and strictly convex in $r$.

\begin{figure}[h]
\caption{First and second derivative of the CVF, as function of the rank
statistic}
\label{fig:CVF_convexity2}{\normalsize \centering
\vspace{5mm} \includegraphics[width=1\linewidth]{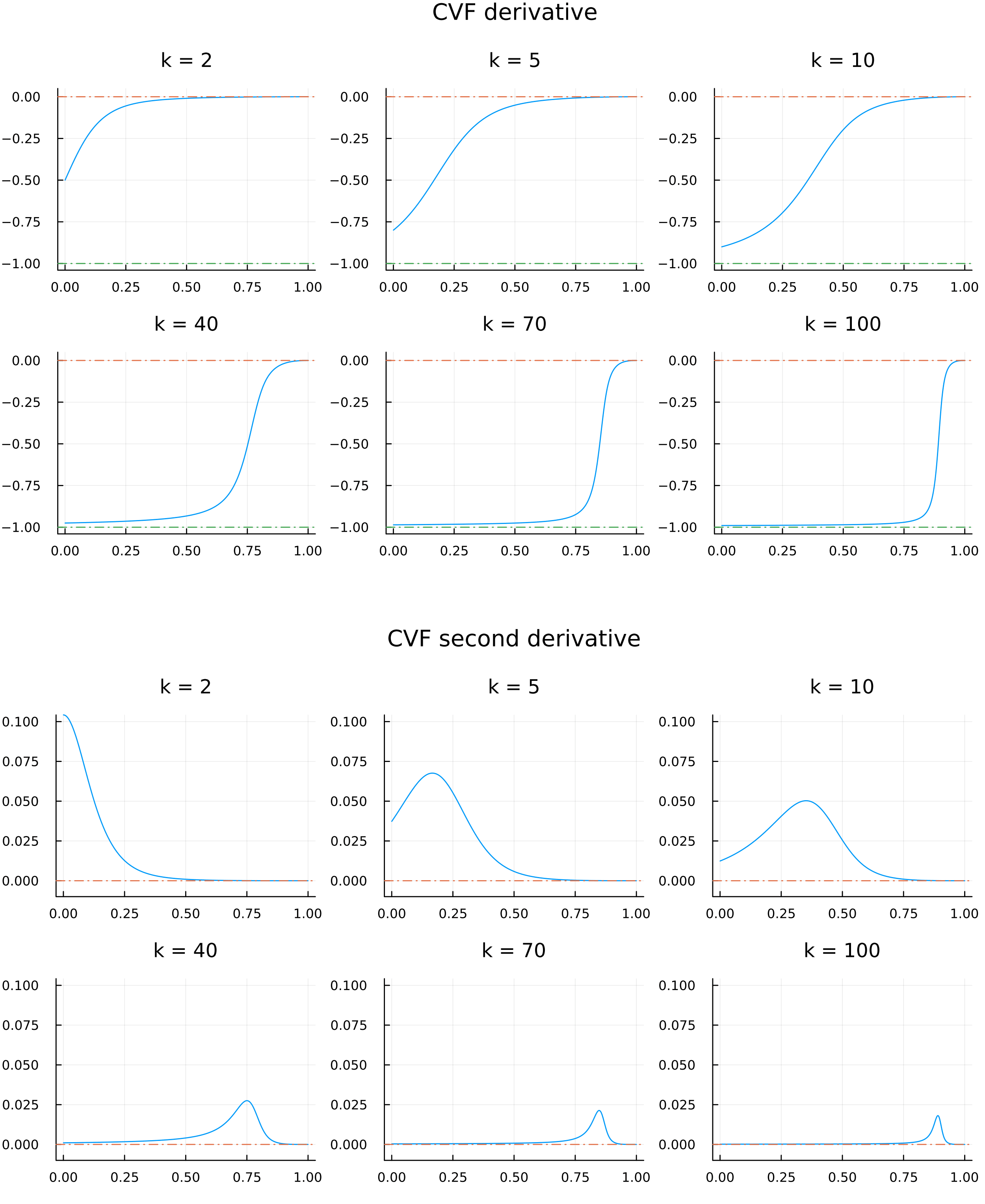} }
\par
{\normalsize {\footnotesize {Note: the domain $[0,+\infty)$ of the rank
statistic is compactified on $[0,1]$ via the transformation $r\mapsto
(2/\pi)\cdot \tan^{-1}(r/20)$.} } }
\end{figure}

Figure~\ref{fig:CVF_convexity2} illustrates these properties for six
representative values of $k$.

\end{document}